\documentclass[aps,onecolumn]{revtex4}
\usepackage{epsfig}
\usepackage{graphicx}
\usepackage{amsmath,amssymb,amsfonts,bm}
\usepackage{array}
\usepackage{url}
\usepackage{hyperref}
\usepackage{multirow}
\usepackage{float}
\usepackage{lineno}
\usepackage{xspace}
\usepackage{ulem}
\usepackage[usenames,dvipsnames]{color}
\usepackage{epstopdf}
\begin{document}
	
	\title{Triple differential cross-section for Laser-assisted (e,2e) process on $H_2O$ molecule by Plane and Twisted electrons Impact}
	\author{Neha}
	\email{p20210062@pilani.bits-pilani.ac.in}

	\author{Rakesh Choubisa}
	\email{rchoubisa@pilani.bits-pilani.ac.in }
	
	\affiliation{Department of Physics,  Birla Institute of Technology and Science, Pilani, Pilani Campus, Vidya Vihar, Pilani, Rajasthan 333031, India}

	\begin{abstract}
	Twisted electron-impact single ionization ((e,2e) process) of water molecule is theoretically studied in the presence of an external laser field. Calculations have been performed in the framework of the first Born approximation in coplanar asymmetric geometry. The wave functions for the fast (incident/scattered) electron and the ejected electron are described by the Volkov and Coulomb Volkov wave functions respectively which take into account the interaction of the laser field with incident/scattered and ejected electrons respectively.
    We calculate the triple differential cross section (TDCS) corresponding to the laser-assisted (e,2e) process for different orientations of the linearly polarised laser field. 
         We describe the molecular state of $H_{2}O$ by the linear combination of atomic orbitals (self-consistent field LCAO method). The angular profile of the TDCS for plane wave electrons is strongly modified by the laser field compared to that of without laser field. However, for the twisted electrons with laser field the angular distribution is slightly changed from those of without laser field. We study the angular distribution of TDCS for laser field polarization parallel to the incident momentum ($\varepsilon_0$ $\parallel$ $k_i$), parallel to momentum transfer ($\varepsilon_0$ $\parallel$ $\Delta$) and perpendicular to the momentum transfer ($\varepsilon_0$ $\perp$ $\Delta$) and we observe that the $\varepsilon_0$ $\parallel$ $\Delta$ has the highest magnitude of TDCS. For the orientations 
            $\varepsilon_0$ $\parallel$ $k_i$ and $\varepsilon_0$ $\perp$ $\Delta$, the laser-assisted plane wave studies shows oscillatory nature of TDCS but for the orientation $\varepsilon_0$ $\parallel$ $\Delta$ we observe only recoil peak for $p-like$ character orbitals and dual peak, a recoil and binary peaks for the $s-like$ character orbital. For laser-assisted twisted electron beam we observe forward and backward peaks for all three orientations of laser-field polarisation.

	\end{abstract}
	\date{\today}
	\maketitle

  \section{Introduction}
  \label{intro}
  Electron collisions on atomic, molecular, and surfaces targets are important probes in modeling and understanding laboratory plasmas, astrophysical processes, laser dynamics, and other fields \cite{bartschat2016electron,christophorou2000electron}. A deep understanding of the electronic structure of atoms and molecules is essential for numerous fields of physics, chemistry, and biology for which electron impact ionization processes offer valuable insights \cite{bartschat2016electron,shalenov2017scattering,de2019relativistic}. The ionization of an atom or molecule by the impact of an electron, also known as the (e,2e) process, is one of the most important collision processes. In an (e,2e) process, the incident electron ionizes the target by ejecting one of the bound electrons and get scattered. The ejected and scattered electrons are detected with their momenta fully resolved \cite{campeanu2018electron}. Complete information on (e,2e) processes is contained in the triple differential cross section (TDCS), which gives the probability of detecting the outgoing electrons with their momenta fully resolved. Significant progress has been made in studying the TDCS for various atomic and molecular targets, supported by experimental and reliable theoretical results \cite{colgan2002time, lahmam2009dynamics,ren2015kinematically}. \\
  In the past few decades, the study of laser-assisted electron collision processes has attracted much attention. The study of laser-assisted electron-atom collisions is highly intriguing from a fundamental perspective. It has potential applications in different fields such as plasma physics, and astrophysics which require scattering cross-section data \cite{bartschat2016electron,christophorou2000electron}. Various investigations of TDCS in different kinematical arrangements give a general understanding of these studies \cite{mittleman2013introduction,francken1990theoretical,ehlotzky1998electron,ehlotzky2001atomic}. Early theoretical studies of laser-assisted ionization processes have been done by neglecting the target dressing effects of the target however they were incorporated in the final state by describing them as Volkov or Coulomb-Volkov states \cite{MOHAN1978399,cavaliere1980particle, mandal1984electron,cavaliere1981effects,banerji1981electron,zangara1982influence,zarcone1983laser}. Later, dressing effects of the target due to the laser field was considered by Joachain and his colleagues \cite{joachain19882,martin1989electron}. Their findings showed significant variations in the differential cross-section. Numerous theoretical studies on  laser-assisted single and double ionization by electron-impact of atomic targets have been reported in the literature \cite{ghosh2009multiphoton,li2007ionization,li2005laser,chattopadhyay2005ionization,van2001double,sanz1999semiclassical,makhoute1999light,taieb1991light,khalil1997laser}. These investigations have also been extended to explore ionization processes in the presence of bi-chromatic laser fields \cite{milovsevic1997electron,ghalim1999electron}. 
  The laser field-free experimental studies have been done on different atomic targets \cite{ehrhardt1986differential,ehrhardt1969ionization,ehrhardt1982triply}. The laser-assisted electron-impact ionization experimental study was conducted on helium \cite{hohr2005electron,hohr2007laser}. These studies have opened new avenues for both theoretical and experimental research in this area. These studies have been reported for the conventional electron beam (plane wave), which doesn't carry any orbital angular momentum. \\ 
	An electron vortex beam (also known as ``twisted electron beam")  carries orbital angular momentum (OAM) along the direction of electron beam propagation \cite{uchida2010generation}. Twisted electron beams (TEB) are characterized by spiraling wavefronts that generate nonzero orbital angular momentum  $(m_{l})$ along their propagation direction (assuming the beam is propagating along the z-axis). 
     These beams carry a helical phase front $e^{im_{l}\phi}$ with the azimuthal angle $\phi$ in the xy-plane (with respect to the x-axis) \cite{lloyd2017electron,bliokh2017theory,larocque2018twisted}.
     
     The theoretical work on non-relativistic electrons significantly advanced experimental and theoretical research on electron states with vortices \cite{bliokh2007semiclassical}. The characteristics of twisted electron beams, such as their transverse momentum, nonzero angular momentum (OAM) along the propagation direction, and helical wavefronts, provide additional degrees of freedom to further study the finer details of the interactions between twisted electrons and atomic or molecular targets. These interactions differ from those observed in conventional untwisted electron beam studies. Twisted electron beams open up opportunities for research in various fields, including optical microscopy, high resolution imaging, quantum state manipulation, quantum computing and information, optical tweezers, advanced information carriers (qudits, qutrits etc.), astronomy, higher-order harmonic generation, etc \cite{verbeeck2010production,mcmorran2011electron,o2002intrinsic,larocque2018twisted,furhapter2005spiral,berkhout2009using,gemsheim2019high}. The opening angle $\theta_p$ and OAM of twisted electron beams influence the ionization processes \cite{van2016electron}. Therefore, to learn about the applications of TEBs in various fields, it is crucial to understand the interaction of electron beams with nonzero OAM and $\theta_p$ at the atomic or molecular scales. So far, theoretical studies have been reported on TEB impact ionization \cite{harris2019ionization,harris2023controlling,dhankhar2020electron,dhankhar2022triple,dhankhar2022twisted,dhankhar2023dynamics,mandal2021semirelativistic}, double ionization, excitation, inelastic scattering, and elastic scattering processes \cite{dhankhar2020double,van2014rutherford,van2015inelastic}.
 Recently, our group also examined the laser-assisted (e,2e) process by the impact of TEB on the hydrogen atoms \cite{Neha2024}. 
    
    For the (e,2e) process on $H_{2}O$ molecule, different theoretical models have been employed to study TDCS for different kinematics. Champion et al. \cite{champion2006single} used various theoretical models to analyze the differential cross sections of (e,2e) processes. 
    The one-Coulomb wave function (1CW) combined with Gaussian-type orbitals (GTO) has also been applied by Champion et al. \cite{champion2009electron} to study (e,2e) on $H_{2}O$. Other models, such as the generalized Sturmain function (GSF), the analytical 1CW model \cite{sahlaoui2011cross}, the two-molecular three-body distorted wave approach (M3DW) \cite{ren2017electron}, the multicentre three distorted waves (MCTDW) have also been used to study such processes \cite{gong2018multicenter}. In addition to this, the second-order distorted wave Born approximation (DWBA2) \cite{singh2019low} has also been explored. Beyond (e,2e) processes, theoretical studies of (e,3e) (double ionization) processes have been reported on $H_2O$ \cite{champion2010molecular,jones2011electron,oubaziz2015h}.
    
    To the best of our knowledge, the study of the (e,2e) process on $H_2O$ molecule has mostly been done for plane waves and very few for TEB without any external laser field. In this paper, for the first time, we attempt to study the laser-assisted (e,2e) process of $H_2O$ molecule with a plane wave and TEB. Our model is based on first-Born approximation. We describe the incident and scattered electron by Volkov wavefunctions, ejected electron by Coulomb Volkov wavefunction, and the molecular states by the linear combination of atomic orbitals (LCAO). 
	The paper is organized as follows:
	In section  \ref{form}, we present our theoretical model employed to study the laser-assisted (e, 2e) process for the plane wave and the twisted electron beam impact.
	In section \ref{result}, we discuss our numerical results for outer orbitals, namely $1b_1$, $1b_2$, $3a_1$, and $2a_1$, of $H_2O$ molecule for different parameters of laser field and twisted electron beam.
	Section \ref{conc} presents our conclusions.
	Throughout the paper, atomic units are used unless stated otherwise.

	\section{\textbf{Theory}}
	\label{form}
	
	\subsection{Preliminaries}
	This section presents the theoretical formalism of laser-assisted (e,2e) process for plane wave and twisted electrons. We chose the water molecule because the water molecule is fundamental to biological processes. The study of the ionization of water molecules has significant applications in radiology, radiation therapy, and planetary atmospheric studies \cite{boudaiffa2000resonant,blanc2015saturn,alizadeh2015biomolecular}. Studying ionization of water molecules with electron impact is crucial to learn about the charged particle interaction in a biological medium.
    
   The basic (e,2e) process on $H_{2}O$ molecule is described as:
    \begin{equation}\label{1}
		e^-(\mathbf{k_i}) + H_{2}O   \rightarrow  H_2O^+ + e^-(\mathbf{k_s}) + e^-(\mathbf{k_e})
	\end{equation}
    Here, a fast-moving electron with incident momentum $\boldsymbol{k_i}$ interacts with the target $H_2O$, results in the emission of a target electron (i.e. a slow moving ejected electron) with momentum $\boldsymbol{k_e}$ and gets scattered with momentum $\boldsymbol{k_s}$ (scattered electron) ($\boldsymbol{k_s}$ $\boldsymbol{>>}$ $\boldsymbol{k_e}$) with all three momenta $\boldsymbol{k_i}$, $\boldsymbol{k_s}$, and $\boldsymbol{k_e}$ are constrained in the same plane. 
    Let us consider the (e,2e) process in the presence of an external laser field as;
    \begin{equation}\label{2}
		e^-(\mathbf{k_i}) + H_{2}O + l\omega  \rightarrow  H_{2}O^+ + e^-(\mathbf{k_s}) + e^-(\mathbf{k_e})
	\end{equation}
    Where positive integer values of $l$ correspond to the absorption of photons, negative integers correspond to the emissions of photons, and $l$ = 0 is related to no photon transfer.
     In this communication, we consider the laser field in a classical framework characterized as a linearly polarized and spatially homogeneous electric field with its magnetic field ignored. The electric field strength experienced by the target electrons is weak and is insufficient to participate in the ionization process. The electric field $\varepsilon(t)$ of the laser is $\boldsymbol{\varepsilon}(t)$  =  $\boldsymbol{\varepsilon_{0}}$ $sin($$\omega$t)$\hat{\varepsilon}$ and the corresponding vector potential $\mathbf{A(t)}$ = $\mathbf{A_{0}}$ $cos$($\omega$t) $\hat{\varepsilon}$ where $\mathbf{A_{0}}$ = c$\mathbf{\varepsilon_{0}}$ $\hat{\varepsilon}$/$\omega$ where, $\varepsilon_{0}$ is the field strength and $\omega$ is the laser frequency, $\hat{\varepsilon}$ is the polarization direction and $c$ is the speed of light.
  
	Here, we compute the TDCS within the framework of the first Born approximation (FBA). The electron impact ionization of the water molecule is treated as a purely electronic transition. In the asymmetric coplanar geometry, because of the large energy difference in the incident/scattered and the bound/ejected electrons, we neglect the exchange effects between the incident/scattered and the bound/ejected electron in the theoretical formalism. Applying the frozen core approximation, the problem of N = 10 electrons can be simplified to a single active electron problem. In the frozen core approximation, only one target electron is assumed to be an active, while the remaining electrons (the passive electrons) remain frozen during the ionization process \cite{de2015theoretical,sahlaoui2012electron}.

In the FBA , the five-fold differential cross section for (e,2e) process is given as:
\begin{equation}\label{3}
    \sigma^5(\alpha,\beta,\gamma) = \frac{d^5\sigma}{d\omega d\Omega_e d\Omega_s dE_e} = (2\pi)^4 \frac{k_e k_s}{k_i} |T_{fi}|^2
\end{equation}
Here, $\alpha$, $\beta$, and $\gamma$ are the Euler angles of the water molecule. $d\Omega$ = $sin\beta$$d\beta$$d\alpha$$d\gamma$ represents the solid angle element for the molecular orientation in the laboratory frame, and $dE_e$ is  ejected electron energy interval. $d\Omega_s$ and $d\Omega_e$ denote the solid angle intervals for the scattered and ejected electrons, respectively. The term $T_{fi}$ is the transition matrix element from the initial state $\psi_i$ to the final state $\psi_f$. For the laser-assisted (e,2e) process the transition matrix element is given as {\cite{joachain19882}}:
	\begin{equation}\label{4}
		T_{fi}^{B1}=-i\int_{-\infty}^{+\infty} dt\left< \chi _{k_{s}}\left ( \mathbf{r_{0}},t \right )\phi_{k_{e}}\left ( \mathbf{r_{1}},t \right ) |V| \chi _{k_{i}}\left ( \mathbf{r_{0}},t \right )\Phi_{0}\left ( \mathbf{r_{1}},t \right )\right>
	\end{equation}
	\begin{equation}\label{5}
		V(r) = \frac{-8}{r_0} - \frac{1}{|\boldsymbol{r_{0}} - \boldsymbol{R_{OH_{1}}}|} - \frac{1}{|\boldsymbol{r_{0}} - \boldsymbol{R_{OH_{2}}}|} + \sum_{i=1}^{10} \frac{1}{|\boldsymbol{r_{0}} - \boldsymbol{r_{i}|}}
	\end{equation}
	
	$V(r)$ describes the Coulomb potential between the projectile and the molecular target. In equation (5), $\mathbf{r_0}$ is the position vector of the incident (and scattered) electron, $\boldsymbol{R}_{OH_{1}}$ and $\boldsymbol{R}_{OH_{2}}$ are the coordinates of the two hydrogen nuclei with respect to the oxygen nucleus, located at the origin of the coordinate system with $|\boldsymbol{R_{OH_{1}}}|$ = $|\boldsymbol{R_{OH_{2}}}|$ = 1.814 a.u.. $\boldsymbol{r_i}$ represents the coordinates of the $i^{th}$ bound electron of the target with respect to the centre of the oxygen nucleus. 
    In the frozen core approximation, the Coulomb potential is approximated to the $V(r')$, where the (Z-1) passive electrons are shielded by nucleus, given as;
    \begin{equation}\label{6}
        V(r') = \frac{-1}{r_0}  +  \frac{1}{|\mathbf{r_0} - \mathbf{r_i}|}
    \end{equation}
 
In the equation (\ref{4}), wave functions $\chi_{k_{i}}$ and $\chi_{k_{s}}$ are the dressed initial and final states of the projectile electrons in the external laser field, described as Volkov wave functions {\cite{joachain19882}} :
	\begin{equation}\label{7}
		\chi_{k_{i,s}}(\mathbf{r_0},t) = (2\pi)^{-3/2}exp[i(\mathbf{{k}_{i,s}}\cdot \mathbf{{r}_{0}} -\mathbf{{k}_{i,s}}\cdot {\boldsymbol{\alpha}_{0}} sin(\omega t) - E_{k_{i,s}} t)]
	\end{equation}
	Where $E_{i,s}$ = $k_{i,s}^2/2$ are the kinetic energies of the projectile in the absence of the leaser field and $\boldsymbol{\alpha_0}$ = $\boldsymbol{\varepsilon_{0}}/ \boldsymbol\omega^2$, $\omega$ is the laser frequency.
    The wave functions
	$\Phi_{0}\left ( \mathbf{r},t \right )$ and $\phi_{k_{e}}\left ( \mathbf{r},t \right )$ in (4) are the dressed states of the target molecule $H_2O$ and the ejected electron respectively. 
  \subsection{Dressed target states}  
  In the present study, we have assumed that the applied laser field is significantly weaker than the Coulomb
binding field in the atom ($\varepsilon_{0}<< (e/a_{0}^{2}\simeq 5\times10^{11} V/m^{-1})$) making it insufficient to ionize the target. Instead, the laser field  is treated perturbatively, which modifies the states of the target, incident, scattered and ejected electrons (dressed states). The initial dressed bound state of the target, $\Phi_{0}\left ( \mathbf {r},t \right )$ is obtained by solving the time-dependent  schr$\ddot{o}$dinger equation using first-order time-dependent perturbation theory and is expressed as: \cite{li1999laser}; 
	\begin{equation}\label{8}
		\Phi_{0}\left (\mathbf{r_{i}},t \right )=exp\left ( -iE_{0}t \right )exp\left ( -i\mathbf{a}\cdot \mathbf{r_{i}}\right )\left [ \Phi_{j}\left ( \mathbf{r_{i}} \right )+\frac{i}{2}\sum_{n}\left [ \frac{exp\left ( i\omega t \right )}{E_{n}-E_{0}+\omega}-\frac{exp\left ( -i\omega t \right )}{E_{n}-E_{0}-\omega} \right ] M_{nj}\Phi_{n}\left ( \mathbf{r_{i}} \right ) \right ] 
	\end{equation}.
	
	Where $\Phi_{j}$ is the ground state wavefunction of the water molecule, expressed as the linear combination of the Slater-type functions (self-consistent field LCAO) centered at the oxygen nucleus \cite{moccia1964one},
 $\mathbf{a}$ = $\mathbf{A}/c$, $\boldsymbol{r_i}$ is the coordinate of the bound electron of the target. Here $exp(-i\mathbf{a} \cdot \mathbf{r_{i}})$ acts as a gauge factor, $\Phi_{n}$ is a state of the target with energy $E_n$ in the absence of the laser field and $ M_{nj}=\left< \Phi_{n} \left|\boldsymbol{\varepsilon_{0}}\cdot \mathbf{r_{i}} \right|\Phi_{j}\right>$. 
 The summation in Eq. (8) runs over the discrete and continuum states of $H_{2}O$ molecule.

 The initial state of $H_{2}O$ comprises ten bonded electrons distributed across five one-centre molecular orbitals, corresponding to the orbitals $1b_1$, $3a_1$  $1b_2$, $2a_1$ and $1a_1$. A dominant atomic orbital component characterizes each molecular orbital in the LCAO framework. The orbital $1b_1$ has $2_{p+1}$, $3a_1$ has $2_{p0}$, $1b_2$ has $2_{p-1}$, $2a_1$ and $1a_1$ has $1s$ dominant atomic orbital character
\cite{champion2001influence}, and expressed as linear combinations of Slater-type functions \cite{moccia1964one}. For the $H_{2}O$ wave function, we use the same mathematical representations as used by Champion et. al. \cite{champion2006single} to describe them here. It is described as;
\begin{equation}\label{9}
    \Phi_{j}(\boldsymbol{r}) = \sum_{k=1}^{N_j}a_{jk} \phi_{n_{jk}l_{jk}m_{jk} }^{\xi_{jk}}
\end{equation}
Where $N_j$ represents the number of Slater functions used to describe the $j-th$ molecular orbital and $n_{jk}$, $l_{jk}$ and $m_{jk}$ are the quantum numbers associated with the $j-th$ molecular orbital. $a_{jk}$ denotes the weight of each atomic component. $\phi_{n_{jk} }^{\xi _{jk}} l_{jk}m_{jk}(\boldsymbol{r})$ is expressed as \cite{champion2005erratum};
\begin{equation}\label{10}
    \phi_{n_{jk}l_{jk}m_{jk}}^{\xi _{jk}} (\boldsymbol{r}) = R_{n_{jk}}^{\xi _{jk}}(r)S_{l_{jk}m_{jk}}(\boldsymbol{\hat{r}})
\end{equation}
Where, $R_{n_{jk}}^{\xi _{jk}}(r)$ the radial part defined by \cite{walker2016electronic}:
\begin{equation}\label{11}
    R^{\xi_{jk}}_{n_{jk}}(r_1) = \frac{(2\xi_{jk})^{n_{jk}+\frac{1}{2}}}{\sqrt{2n_{jk}!}}r_1^{n_{jk}-1}e^{-\xi_{jk}r_1},
\end{equation}
$S_{l_{jk}m_{jk}}(\boldsymbol{\hat{r}_1})$ is the real spherical harmonics given as for $m_{jk} \neq 0$,
\begin{equation}\label{12}
\begin{aligned}
S_{l_{jk}m_{jk}}(\boldsymbol{\hat{r}}_1) = {} & \sqrt{\Bigg( \frac{m_{jk}}{2|m_{jk}|} \Bigg)} \Bigg\{  Y_{l_{jk}-|m_{jk}|}(\boldsymbol{\hat{r}}_1)+ \\
&  (-1)^m \Bigg( \frac{m_{jk}}{|m_{jk}|} \Bigg)Y_{l_{jk}|m_{jk}|}(\boldsymbol{\hat{r}}_1)  \Bigg\},
\end{aligned}
\end{equation}
and for $m_{jk} = 0$:
\begin{equation}\label{13}
     S_{l_{jk}0}(\boldsymbol{\hat{r}}_1) = Y_{l_{jk}0}(\boldsymbol{\hat{r}}_1).
\end{equation}
 
Here $Y_{lm}$ represents the complex spherical harmonics.

	The Coulomb-Volkov wave function is expressed as \cite{joachain19882}:
	\begin{equation}\label{14}
		\begin{aligned}
			\phi_{k_{e}}\left ( \mathbf{r_{i}},t \right )=exp\left ( -iE_{k_{e}} t \right )exp\left ( -i\mathbf{a}\cdot \mathbf{r_{i}} \right )exp\left ( -i\mathbf{k_{e}} \cdot \boldsymbol{\alpha_{0}} sin\omega t  \right ) \times \\ \left [ \psi_{C,k_{e}}^{-}\left ( \mathbf{r_{i}} \right )+ \frac{i}{2}\sum_{n}\left [ \frac{exp\left (i\omega t  \right )}{E_{n}-E_{k_{e}}+\omega}- \frac{exp\left (-i\omega t  \right )}{E_{n}-E_{k_{e}}-\omega}\right ]M_{n,k_{e}}\psi_{n}\left ( \mathbf{r_{i}} \right )+i\mathbf{k_{e}}\cdot \boldsymbol{\alpha _{0}} sin(\omega t) \psi_{C,k_{e}}^{-}\left ( \mathbf{r_{i}} \right )\right ];
		\end{aligned}
	\end{equation} 
	Where $\psi_{C,k_{e}}^{-}( r_{1})$ is the Coulomb wave function, expressed as;
	\begin{equation}\label{15}
		\psi_{c,k_{e}}^{-}=\left (2\pi   \right )^{-3/2}e^{\pi/2\mathbf{k_{e}}}e^{i\mathbf{k_e} \cdot \mathbf{r}} \Gamma\left ( 1+i/k_{e}\right )_1F_{1}\left [ -i/k_e,1,-i\left (k_{e}r_{1}+\mathbf{k_{e}}\cdot \mathbf{r} \right ) \right ],
	\end{equation}
	where $E_{k_{e}}$ is ejected electron energy, $M_{n,k_{e}}=\left< \Phi_{n} \left|\boldsymbol{\varepsilon_{0}}\cdot \mathbf{r} \right|\psi_{C,k_{e}}^{-}\right>$.
	Using equations (\eqref{6}) - (\eqref{15}) in the first Born $\mathit{T}$-matrix element (\eqref{4}) and integrating over time, we find;
	\begin{equation}\label{16}
		T_{fi}^{B1}=\left ( 2\pi  \right )^{-1} i\sum_{l=-\infty}^{+\infty}\delta \left (E_{k_{s}} + E_{k_{e}} - E_{k_{i}}- E_{0}-l\omega  \right )f_{ion}^{B1,l}
	\end{equation}
	Here $f_{ion}^{B1,l}$ involves the transfer of $l$ photons, and is given as {\cite{joachain19882}};
	\begin{equation}\label{17}
		f_{ion}^{B1,l}=f_I+f_{II}+f_{III}
	\end{equation}
	where, 
	\begin{equation}\label{18}
		f_{I}=-2\mathbf{\Delta^{-2}} J_{l}\left ( \lambda  \right ) \left<\psi_{C,k_{e}}^{-}\left|exp\left ( i\mathbf{\Delta} .\mathbf{r} - 1 \right ) \right|\Phi_{j} \right>,
	\end{equation}\\
	\begin{equation}\label{19}
		f_{II}=i\mathbf{\Delta^{-2}}\sum_{n}\left< \psi_{C,k_{e}}^{-}\left| \exp(i\mathbf{\Delta} .\mathbf{r} - 1)\right|\Phi_{n}\right>M_{nj}\left [ \frac{J_{l-1}\left ( \lambda  \right )}{E_{n}-E_{j}-\omega } -\frac{J_{l+1}\left ( \lambda  \right )}{E_{n}-E_{j}+\omega } \right ]
	\end{equation}\\
	\begin{equation}\label{20}
		\begin{aligned}
			f_{III}=i\mathbf{\Delta^{-2}}\sum_{n}\left< \Phi_{n}\left| \exp\left ( i\mathbf{\Delta} .\mathbf{r} - 1\right )\right|\Phi_{j}\right>M_{n,k_{e}}^{*}\left [ \frac{J_{l-1}(\lambda )}{E_{n}-E_{k_{e}}+\omega }-\frac{J_{l+1}\left ( \lambda  \right )}{E_{n}-E_{k_{e}}-\omega } \right ]\\
			-\mathbf{\Delta^{-2}}\mathbf{k_{e}}.\mathbf{\alpha_0}\left [J_{l-1} \left (\lambda   \right )-J_{l+1}\left ( \lambda  \right )  \right ]\left< \psi_{C,k_{e}}^{-}\left| exp\left ( i\mathbf{\Delta} .\mathbf{r} \right )\right|\Phi_{j}\right>
		\end{aligned}
	\end{equation}
	
	Here, $J_{l}$ is the Bessel function of order $l$,  $\lambda$ = $(\mathbf{\Delta}- \mathbf{k_{e}})\cdot \boldsymbol{\alpha_{0}}$ where $\mathbf{\Delta}$ = ($\mathbf{k_{i}} - \mathbf{k_{s}}$) is the momentum transfer of collision. $\Phi_j$ is the ground state of target with energy $E_j$ and $\Phi_n$ is the target state of energy $E_n$ in the absence of laser field.
   
    For the absorption of one photon $l$ = 1, the scattering amplitude is derived directly from the expressions (\ref{18}) - (\ref{20}). With the application of the closure property ( ($\sum_{n}|\Phi_n><\Phi_n|$) = 1 the summation over all 'n' states converges to unity) in the expression (\ref{19}) - (\ref{20}) and by keeping only the first order term of the field strength (as it is a weak field) in the expansion of the Bessel function the scattering amplitude for $l$ = 1 can be expressed as;
    \begin{equation}\label{21}
        f_{I}= -\mathbf{\Delta^{-2}} \omega^{-2} \Delta.\varepsilon_0 \left<\psi_{C,k_{e}}^{-}\left|exp\left ( i\mathbf{\Delta} .\mathbf{r} - 1 \right ) \right|\Phi_{j} \right>,
    \end{equation}
    \begin{equation}\label{22}
		f_{II}= -i\mathbf{\Delta^{-2}} \left< \psi_{C,k_{e}}^{-}\left| \exp(i\mathbf{\Delta} .\mathbf{r} - 1)G_{c}(E_j + \omega) \boldsymbol{\varepsilon . r}\right|\Phi_{j}\right>
	\end{equation}\\
    \begin{equation}\label{23}
		\begin{aligned}
			f_{III}=i\mathbf{\Delta^{-2}}\left< \psi_{C,k_{e}}^{-}\left| \boldsymbol{\varepsilon . r} \exp\left ( i\mathbf{\Delta} .\mathbf{r} - 1\right)   G_{c}(E_{k_e} - \omega)\right |\Phi_{j}\right>
		\end{aligned}
	\end{equation}
    Here $G_{c}(E)$ is the Coulomb Green's function.\\
    During gas-phase ionization experiments, the molecules are randomly aligned and aligning them in a specific direction is difficult to achieve. To mimic this scenario in theoretical calculations, we compute TDCS by averaging $\sigma^{5}(\alpha,\beta,\gamma)$ over all possible orientations of the water molecules. The TDCS obtained by integrating the 5DCS over Euler angles, is given as;
    \begin{equation}\label{24}
\frac{d^3\sigma}{d\Omega_e d\Omega_s dE_e} = \frac{1}{8 \pi^2} \int \sigma^{(5)}(\alpha,\beta,\gamma)  \sin \beta d\alpha d\beta d\gamma,
\end{equation}
  The resulting TDCS is given by \cite{sahlaoui2011cross}:
    \begin{equation}\label{25}
\frac{d^3\sigma}{d\Omega_e d\Omega_s dE_e} = \frac{k_e k_s}{k_i} \sum_{k = 1} ^{N_j} \frac{a_{jk}^2}{\hat{l}_{jk}}\sum_{\mu = -l_{jk}}^{l_{jk}} |f_{ion}^{B_{1},l}(\mathbf{\Delta})|^2.
\end{equation}

	\subsection{Twisted beam ionization}
	\label{TEB}
	
This section briefly discusses the theoretical formalism for the laser-assisted (e,2e) process by twisted electron beam. The formalism for the (e,2e) process with a TEB follows a similar framework to that of the plane wave, with the key difference being that, as the projectile, the twisted electron beam replaces the plane wave. The twisted electron beam is a freely propagating beam with a phase singularity at the centre. It carries helical wavefronts with varying phase in the xy-plane, giving orbital angular momentum (OAM) $m_l$ along the propagation direction (z-axis). The twisted electron beam has a cylindrical symmetry therefore one can use the Bessel wave function to describe the twisted electrons. In the cylindrical coordinates system, the twisted electrons are described as \cite{harris2021single}:
\begin{equation}\label{26}
   \psi_{\varkappa m_l}^{\left ( tw \right )}\left ( \mathbf{r_{0}} \right ) = \frac{e^{im_l\phi}}{2\pi}J_{l}(k_{i\perp},r_0)e^{ik_{iz}z}
\end{equation}
Unlike the plane waves, twisted waves have both the longitudinal ($k_{iz}$) and the transverse components ($k_{i\perp}$) of the momentum. These components can be written in terms of the opening angle as \cite{Plumadore2021};  
	\begin{equation}\label{27}
		\mathbf{k}_{i\perp} = (k_i\sin{\theta_p}\cos{\phi_p})\hat{x}+(k_i\sin{\theta_p}\sin{\phi_p})\hat{y}
        \end{equation}
        \begin{equation}\label{28}
            \mathbf{k_{iz}} = k_i\cos{\theta_p}\hat{z}
        \end{equation}

	Here, $\theta_{p}$ and $\phi_{p}$ are the polar and azimuthal angles of the $\mathbf{k_i}$, respectively. The longitudinal component ${k_{iz}}$ of the incident momentum $\mathbf{k_{i}}$ is fixed but the transverse component ($k_{i\perp}$) changes its direction with $\phi_p$. 
   An impact parameter $\textbf{b}$ descries the extent of transverse orientation of target from the incident beam axis. The Bessel wave function or the twisted electron wave function with an arbitrary impact parameter  $\textbf{b}$ can be effectively expressed as a superposition of tilted plane waves (\cite{serbo2015scattering}):
	
	\begin{equation}\label{29}
		\psi_{\varkappa m_l}^{\left ( tw \right )}\left ( \mathbf{r_{0}} \right )= \int_{0}^{\infty }\frac{dk_{i\perp}}{2\pi }k_{i\perp}\int_{0}^{2\pi}\frac{d\phi_p}{2\pi}a_{\varkappa m_l}\left ( k_{i\perp} \right )e^{i \bf{k_{i}.r_0}} e^{-i\bf {k_{i}}.\bf{b}}
	\end{equation}
    \begin{equation}\label{30}
	a_{\varkappa m_l}\left ( k_{i\perp} \right ) = (-i)^{m_{l}} e^{im_l\phi_{p}}\sqrt{{2\pi}} \delta(|{k_{i\perp}}| - \varkappa)
 \end{equation}
	where $a_{\varkappa m_l}\left ( k_{i\perp} \right )$ is the magnitude of the transverse momentum ($k_i$ $sin{\theta_p}$). 
 For the laser-assisted process, straightforwardly, the Bessel beam $\psi_{\varkappa m_l}^{\left ( tw \right )}\left ( \mathbf{r_{0}} \right )$ can be expressed as the superposition of Volkov wave functions \cite{karlovets2012electron};
 
 \begin{equation}\label{31}
		\psi_{\varkappa m_l}^{\left ( tw \right )}\left ( \mathbf{r_{0}} \right )= \int_{0}^{\infty }\frac{dk_{i\perp}}{2\pi }k_{i\perp}\int_{0}^{2\pi}\frac{d\phi_p}{2\pi}a_{\varkappa m_l}\left ( k_{i\perp} \right )e^{i \bf{k_{i}.r_0 -\mathbf{{k}_{i,s}}\cdot {\boldsymbol{\alpha}_{0}} sin(\omega t) - E_{k_{i,s}} t}}e^{-i\bf {k_{i}}.\bf{b}}
	\end{equation}
	
	For the computation of TDCS for laser-assisted (e,2e) process by TEB, we replace the plane wave description ($\chi_{k_i}$) with the Bessel beam $\psi_{\varkappa m_l}^{\left ( tw \right )}$ in equation (\ref{4}). Thus the transition matrix of TEB $T_{fi}^{tw}(\varkappa, \mathbf{\Delta})$ is expressed in terms of the plane wave transition amplitude $T_{fi}^{B1}(\mathbf{\Delta})$  \cite{dhankhar2020double} as:
	\begin{equation}\label{32}
		T_{fi}^{tw}\left ( \varkappa,{\mathbf{\Delta_{tb}}},\mathbf{b} \right )= \left ( -i \right )^{m_l}\sqrt{\frac{1}{2\pi}}\int_{0}^{2\pi}\frac{d\phi_{p}}{2\pi}e^{im\phi_{p}-ik_{i\perp} \cdot {b}}T_{fi}\left ( \mathbf{\Delta} \right ),
	\end{equation}
    where $\mathbf{k_{i\perp}} \cdot \mathbf{b} $ = $\mathbf{\varkappa}$ $b$ $cos(\phi_{p} - \phi_{b})$.

    Note that for the incident twisted electron beam, the momentum transfer can be described as \cite{Plumadore2021};
	\begin{equation}\label{33}
		\Delta_{tb}^{2} = k_{i}^2 + k_{s}^2 - 2 k_{iz} k_{sz} - 2k_{i\perp}k_{s\perp} cos(\phi_p -\phi_s),
	\end{equation}
	where $\phi_{s}$ = 0 for coplanar geometry.
	


In an experimental study with the TEB, it is difficult to precisely put the target at any specific impact parameter $\textbf{b}$. For practical purposes, it becomes crucial to study the macroscopic target to account for the broad range of all possible impact parameters and make the model more realistic.
     The average TDCS for macroscopic targets is calculated by integrating $T_{fi}^{tw}(\Delta)$ over impact parameter, in the transverse plane.
     The average cross-section $(TDCS)_{av}$ is be described as \cite{harris2019ionization} :
	\begin{equation}\label{34}
		(TDCS)_{av}=\frac{1}{2\pi\cos\theta_p}\int^{2\pi}_{0}d\phi_p \frac{d^3\sigma(\mathbf{\Delta})}{d\Omega_{e}d\Omega_{s}dE_{e}}.  
	\end{equation}

	\section{Results and Discussions}
	\label{result}

	 We presents the results of TDCS for the laser-assisted (e,2e) process on ($H_2O$) molecule by the PW and TEB in a coplanar asymmetric geometry. 
    Nikita et al. \cite{dhankhar2022triple} have benchmarked the theoretical results for the ionization of outer orbitals $1b_1$, $3a_1$, $1b_2$, and $2a_1$ of $H_2O$ molecules with experimental results for the plane wave. We extended this model by incorporating the effects of an external laser-field by describing the incident, bound, scattered and ejected electrons by the dressed states (see previous section for more details).

    In our work, the field strength is weak and held constant at $\varepsilon_{o}$  = $10^7$ V/cm, corresponding to the laser intensity 1.32 × $10^{11}$ W $cm^{-2}$. We present the angular profile of TDCS of the ejected electrons for $l = 1$, i.e, absorption of a photon during the collisions, in the coplanar asymmetric geometry. Furthermore, we investigate the influence of various laser and TEB parameters on the angular profile of the TDCS.
    The kinematics used here is incident electron energy $E_i$ = 250eV, the ejected electron energy $E_e$ = 10eV (and 8eV for the orbital $3a_1$), and the scattering angle $\theta_s$ = 15$^\circ$. We compare our results of the PW with the LA-PW for different orientations of the laser-filed vector, namely \boldsymbol{$\varepsilon_0$ $\parallel$ $k_i$}, \boldsymbol{$\varepsilon_0$ $\parallel$ $\Delta$} and \boldsymbol{$\varepsilon_0$ $\perp$ $\Delta$} and LA-TEB for different values of the OAM $m_{l}$ = 1, 2, and 3. 
\subsection{Angular profile of TDCS for Laser-assisted (e,2e) by plane wave}

	\begin{figure*}[htp!]
		\begin{tabular}{cc}
			\includegraphics[width=8.00cm]{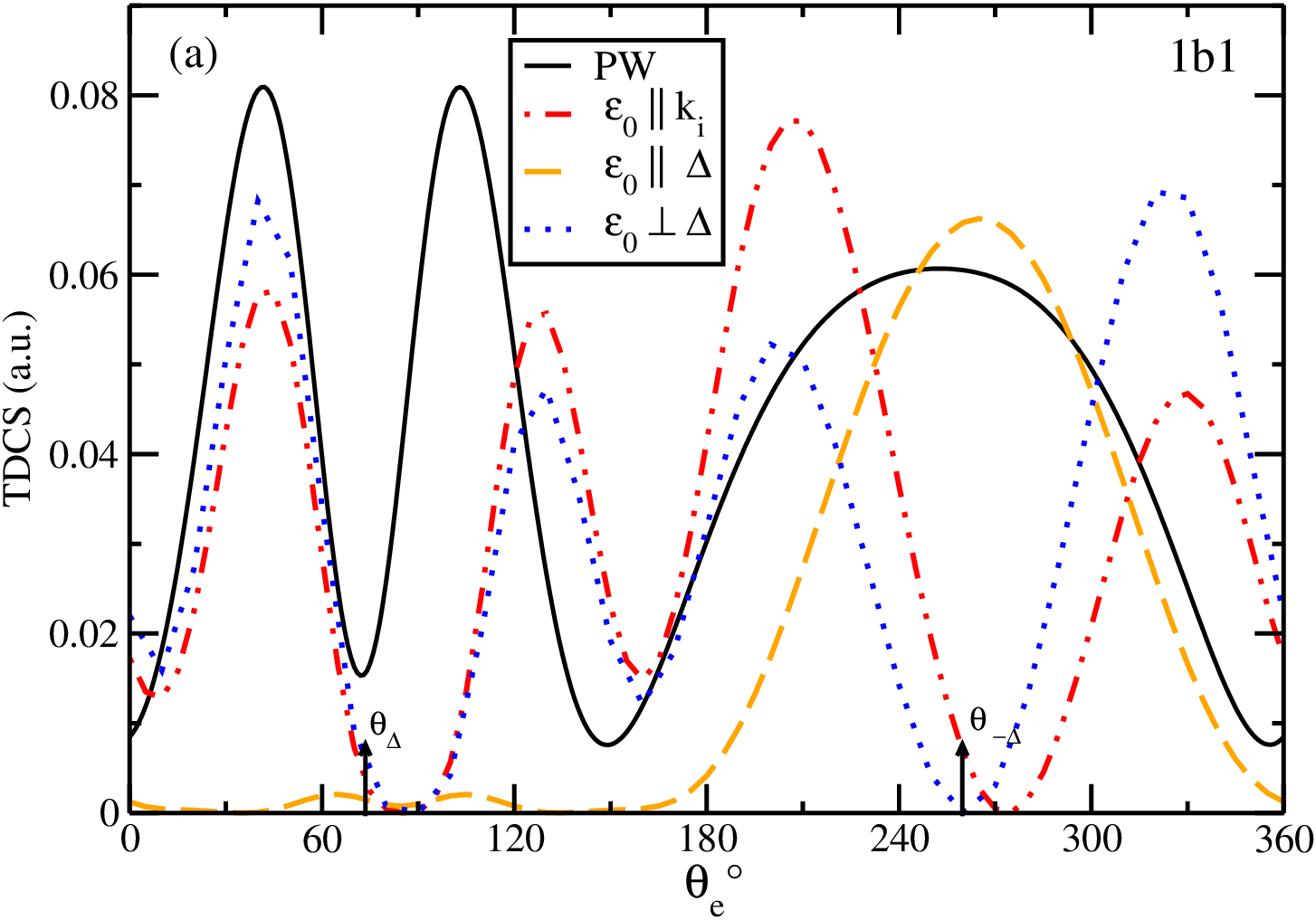}\ &
			\includegraphics[width=8.00cm]{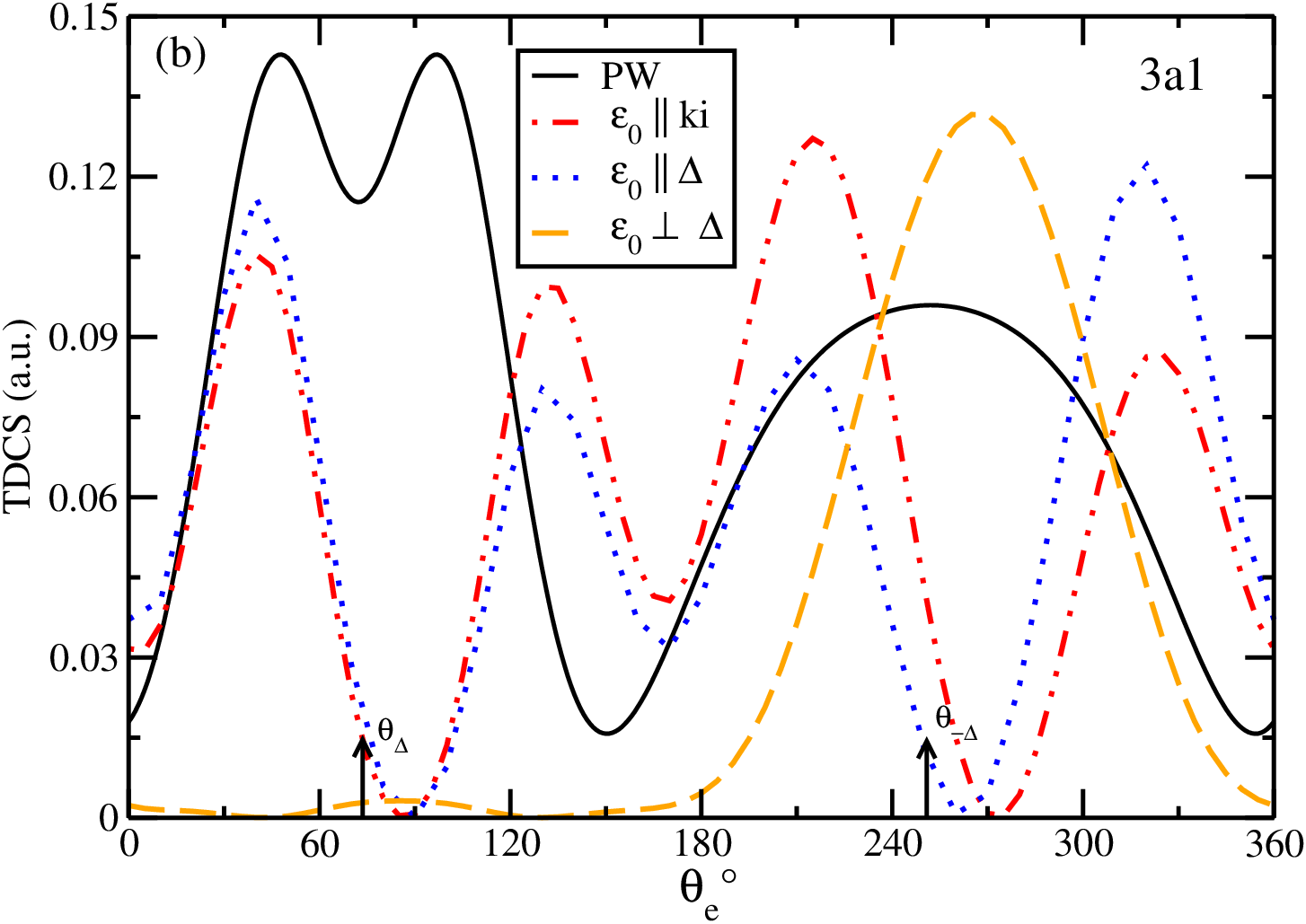}
		\end{tabular}

		\begin{tabular}{cc}
			\includegraphics[width=8.00cm]{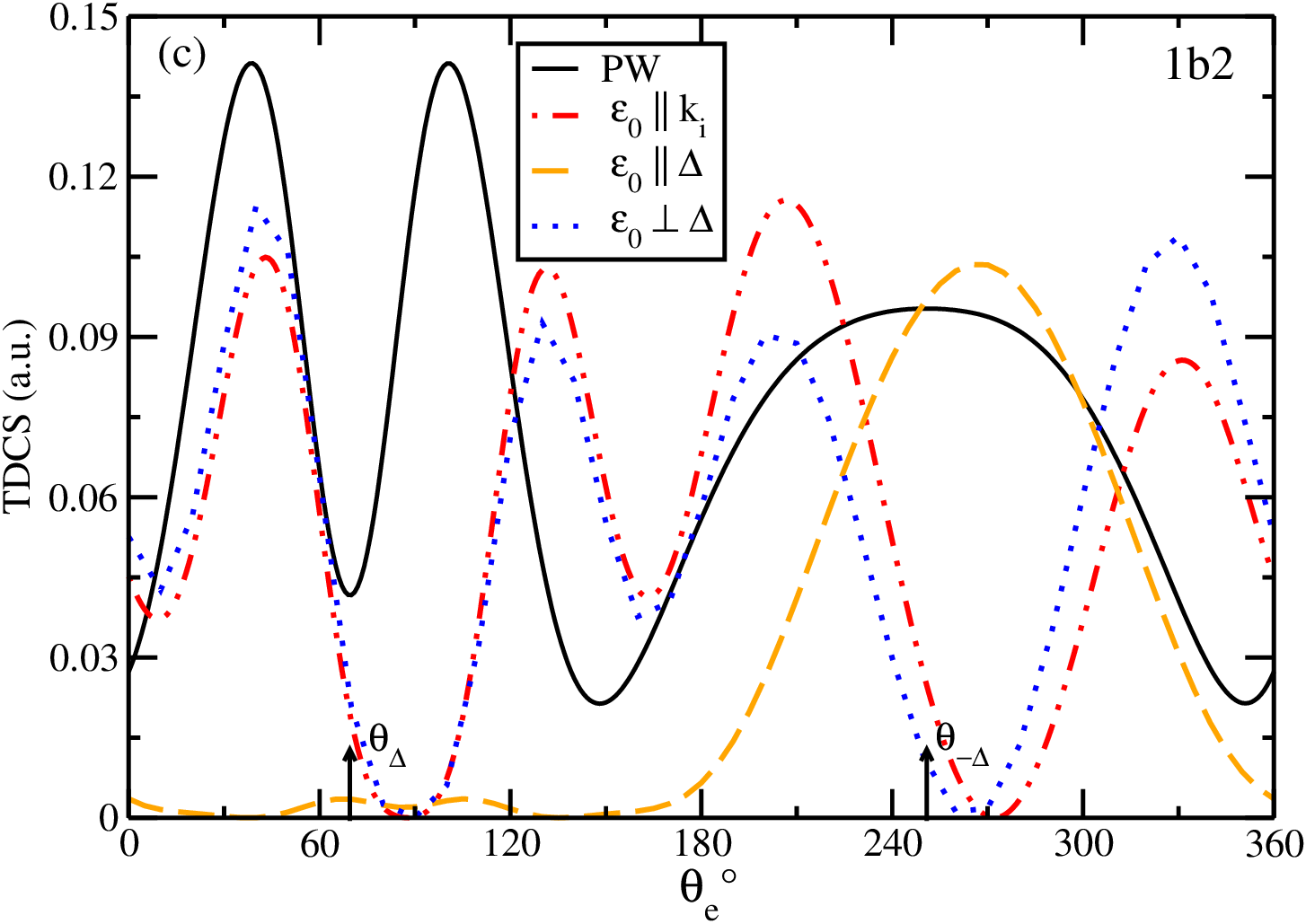}\ & \includegraphics[width=8.00cm]{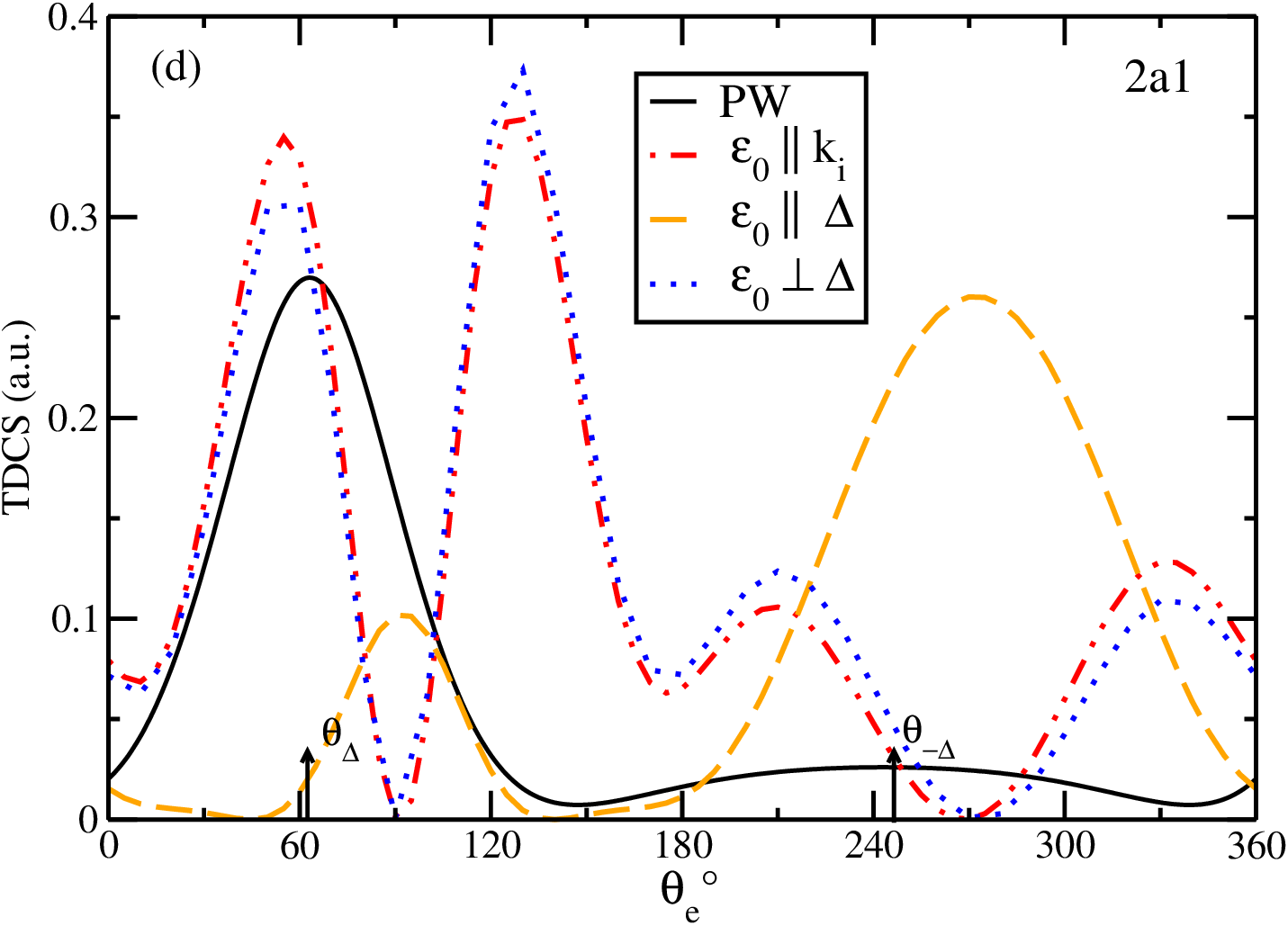}
		\end{tabular}
		\caption{TDCS as a function of $\theta_e$ for laser-assisted (e,2e) process on $H_2O$ molecule by plane wave. The kinematics applied here is $E_i$ = 250eV, $E_e$ = 10 eV, $\varepsilon_{0}$ = $10^7$V/cm, $\hbar$ $\omega$ = 1.17 eV (Nd: YAG laser), and $\theta_{s}$ = 15$^\circ$. 
        TDCS is plotted for plane wave (PW) without laser field solid curve and for the laser-assisted plane wave (LA-PW) at different orientations of laser-field vector ($\varepsilon_0$), $\mathbf{\varepsilon_0}$ $\parallel$ $k_i$ dashed-dotted-dotted curve,  $\mathbf{\varepsilon_0}$ $\parallel$ $\Delta$ dashed curve, and  $\mathbf{\varepsilon_0}$ $\perp$ $\Delta$ dotted curve. The TDCS magnitude for $\mathbf{\varepsilon_0}$ $\parallel$ $k_i$ is scaled up by 100 in figure(a)-(d), and for $\mathbf{\varepsilon_0}$ $\parallel$ $\Delta$ scaled up 20 for orbitals $1b_1$ and $1b_2$, by 15 for orbital $3a_1$ and by 200 for orbital $2a_1$. For $\mathbf{\varepsilon_0}$ $\perp$ $\Delta$ orientation TDCS is scaled up by 100 for orbitals $3a_1$, $1b_1$ and $1b_2$, and $2a_1$ by 500.}  
		\label{fig:1}
	\end{figure*}

	In Fig \ref{fig:1}, we present the TDCS for the laser-assisted (e,2e) process for plane wave electrons as a function of ejected electron angle $\theta_e$, with the exchange of one photon $l$ = 1. As mentioned earlier, the molecular wave function for the various orbitals of the water molecule is formulated using a linear combination of atomic orbitals (LCAO). Where the character of the $1b_1$ orbital is primarily determined by the $2_{p+1}$ atomic orbital, the $3a_1$ by $2_{p0}$ the $1b_2$ by $2_{p-1}$ and the $2a_1$ by $1s$ \cite{champion2001influence,hanssen19942e}. Therefore, the overall behaviour of the TDCS is determined by the dominant atomic component of each molecular orbital. For the validation of our theoretical model, we have benchmarked our results for the plane wave with and without laser field for hydrogen atom \cite{Neha2024}, which are in agreement with the published results \cite{joachain19882}.
    We reproduced the results of Nikita et al. \cite{dhankhar2022triple} to benchmark our theoretical model for the twisted electron beam without laser field. 
       
        In Figs \ref{fig:1}(a)-(c) for orbitals $1b_1$, $3a_1$, and $1b_2$ for PW without laser field (black solid curve), we observed splitted two peak structure in the binary region with dip at $\theta_{\Delta}$ and a single peak, peaked around $\theta_{-\Delta}$ direction. This is due to the dominating $p-like$ character of these orbitals (see Figs \ref{fig:1} (a), (b) and (c)). However, for $2a_1$ orbitals, we observed a binary peak in the direction $\theta_{\Delta}$ and a recoil peak in the $\theta_{-\Delta}$ direction which is due to $s$-type character of the orbital (see Fig \ref{fig:1}(d)) \cite{de2015theoretical,champion2001influence}. The angular distribution of TDCS differs in the laser-assisted (e,2e) process for, $\varepsilon_0$ $\parallel$ $k_i$ (see dashed-dotted-dotted curve Figs \ref{fig:1}(a)-(d)), $\varepsilon_0$ $\parallel$ $\Delta$ (see blue dotted curve Figs \ref{fig:1}(a)-(d)) and $\varepsilon_0$ $\perp$ $\Delta$ (see orange dashed-dashed curve Figs \ref{fig:1}(a)-(d)). For $\varepsilon_0$ $\parallel$ $k_i$, we observed the oscillatory nature of TDCS (dashed-dotted-dotted curve multiplied by 100 to compare with {PW} results). For the $p-like$ orbitals $1b_1$, $3a_1$ and $1b_2$, we observed that recoil peak and binary peak split into lobs of different amplitudes, with four maxima around $\theta_e$ = 40$^\circ$, 135$^\circ$, 215$^\circ$ and 325$^\circ$ (see red dashed-dotted-dotted curve in Figs \ref{fig:1}(a)-(c)). But for the orbital $2a_1$ with $s-like$ character, we observed that binary peak (near $\theta_e$ = 60$^\circ$) splits into two peaks at $\theta_e$ = 55$^\circ$ and 130$^\circ$ and in the region of recoil peak (near $\theta_e$ = 247$^\circ$) we observed minima at $\theta_e$ = 270$^\circ$. Further, two smaller peaks at $\theta_e$ = 210$^\circ$ and 330$^\circ$ are observed for $2a_1$ orbital (see the red dashed-dotted-dotted curve in Figs \ref{fig:1}(d). For the laser-field orientation $\varepsilon_0$ $\perp$ $\Delta$; we observed that the angular distribution of TDCS for $\varepsilon_0$ $\perp$ $\Delta$ is similar to $\varepsilon_0$ $\parallel$ $k_i$ with small shifts in the peak positions for $p-like$ character orbitals (see blue dotted and red dashed-dotted-dotted curves near $\theta_e$ = 240$^\circ$ in Figs \ref{fig:1}(a)-(c)). We further observed that for orientation $\varepsilon_0$ $\perp$ $\Delta$
        the peaks in the direction $\theta_e$ = 30$^\circ$ and 330$^\circ$ direction gets enhanced compared to that for $\varepsilon_0$ $\parallel$ $k_i$ orientation. However the peaks in the direction around $\theta_e$ = 130$^\circ$ and 200$^\circ$, enhanced for $\varepsilon_0$ $\parallel$ $k_i$ orientation and slightly suppressed for $\varepsilon_0$ $\perp$ $\Delta$ orientation (see blue dotted and red dashed-dotted-dotted curves in Figs \ref{fig:1}(a)-(c)). We further observe that there are no significant changes in the peaks for $2a_1$ orbital of $s-like $ character. Unlike the $p-like$ character orbitals, for $2a_1$ orbital the peaks for orientation $\varepsilon_0$ $\perp$ $\Delta$ get enhanced in the direction $\theta_e$ = 150$^\circ$ and 210$^\circ$  and suppressed in the direction $\theta_e$ = 60$^\circ$ and 330$^\circ$ when compared to that for the orientation $\varepsilon_0$ $\parallel$ $k_i$ (see blue dotted and red dashed-dotted-dotted curves in Fig \ref{fig:1}(d)).
                When we consider the orientation $\varepsilon_0$ $\parallel$ $\Delta$, unlike the other two orientations ($\varepsilon_0$ $\parallel$ $k_i$ and $\varepsilon_0$ $\perp$ $\Delta$), we observed a dominant one peak structure (see an orange dashed curve in Fig \ref{fig:1}(a)-(d)). For $p-like$ character orbitals $1b_1$ (TDCS magnitude multiplied by 20 to compare with PW), $3a_1$  (TDCS magnitude multiplied by 15 to compare with PW), and $1b_2$ (TDCS magnitude multiplied by 20 to compare with PW) we observed a recoil peak at $\theta_e$ = 270$^\circ$ and no binary peak. But for orbital $2a_1$, we observed two peaks, a dominant recoil peak at $\theta_e$ = 90$^\circ$ and a shallow binary peak at 270$^\circ$ (TDCS magnitude multiplies by 200 to compare with PW). And out of the three orientations of the laser field, the magnitude of TDCS is maximum for the orientation $\varepsilon_0$ $\parallel$ $\Delta$.

	\subsection{Angular profile of TDCS for the laser-assisted (e,2e) process by twisted electron beam} 
   In this section, we present the TDCS results for the laser-assisted (e,2e) process by the twisted electron beam for $m_l$ = 1 (black dashed curve), 2 (red dashed-dotted-dotted curve), and 3 (blue dashed-dotted curve) for the same kinematics used in the figure \ref{fig:1} with $\theta_s$ = $\theta_p$ = 15$^\circ$. 
			
		\begin{figure*}[htp!]

				\begin{tabular}{cc}
					\includegraphics[width=8.00cm]{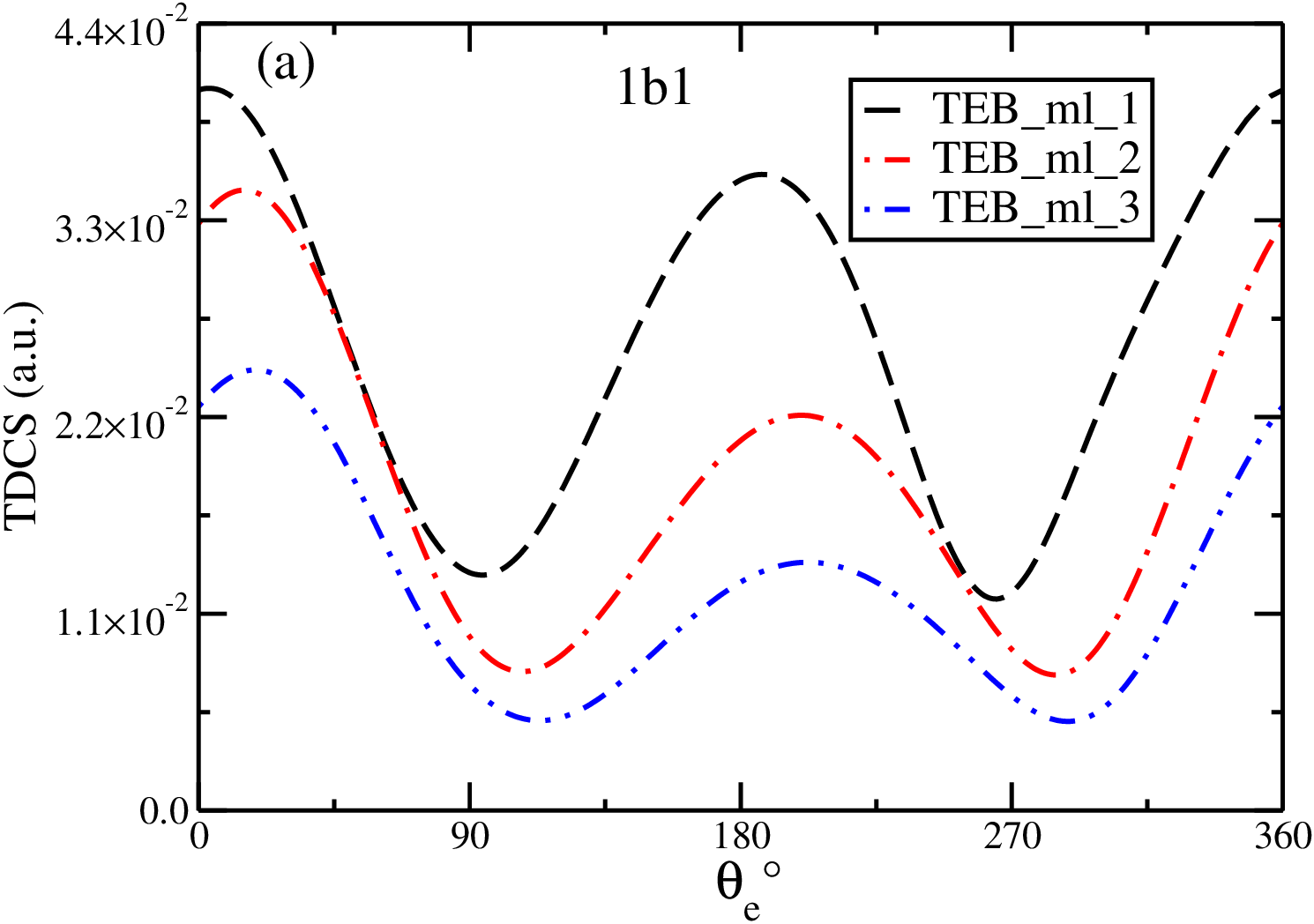}\ & \includegraphics[width=8.00cm]{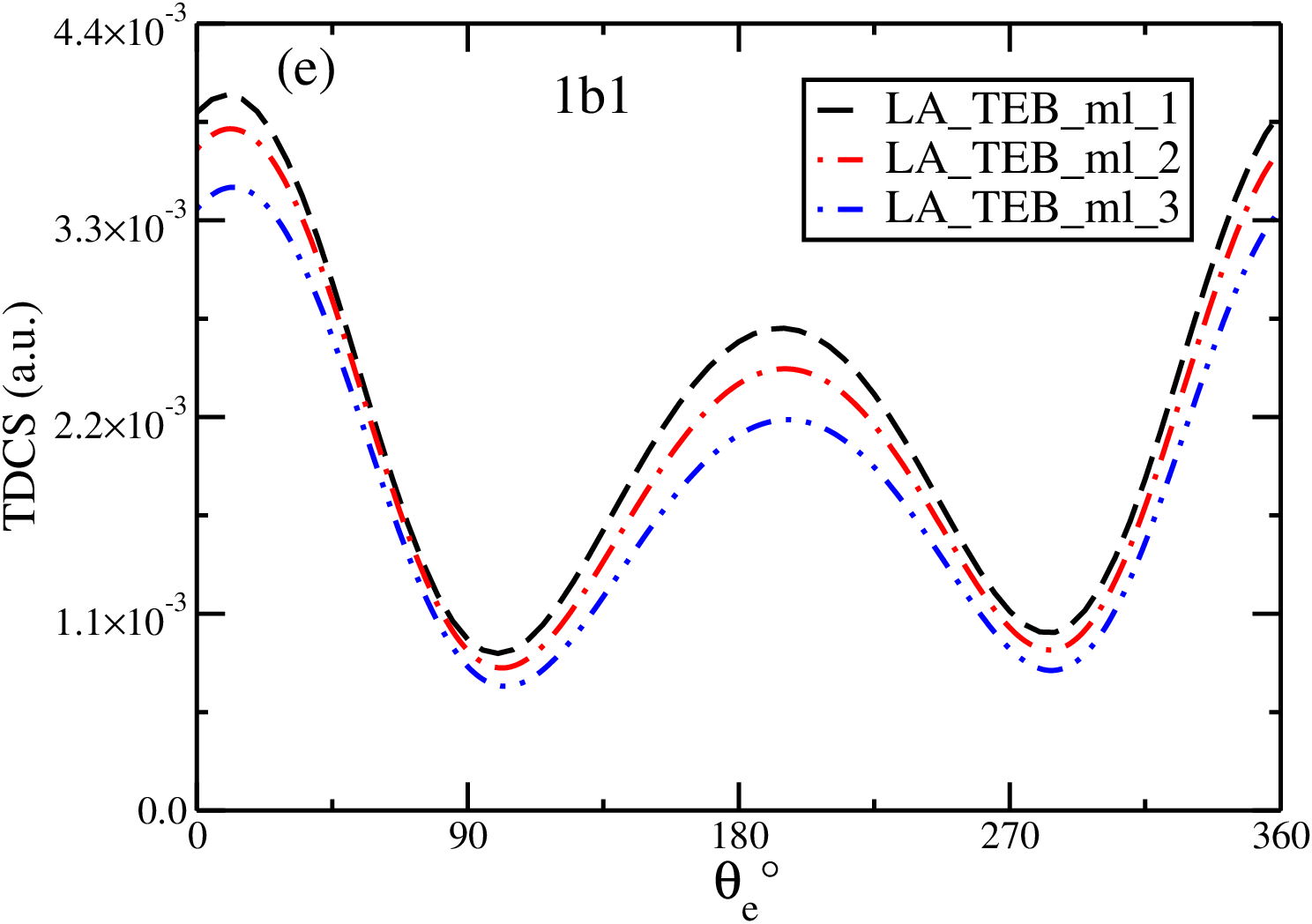}

				\end{tabular}
				
				\begin{tabular}{cc}
					\includegraphics[width=8.00cm]{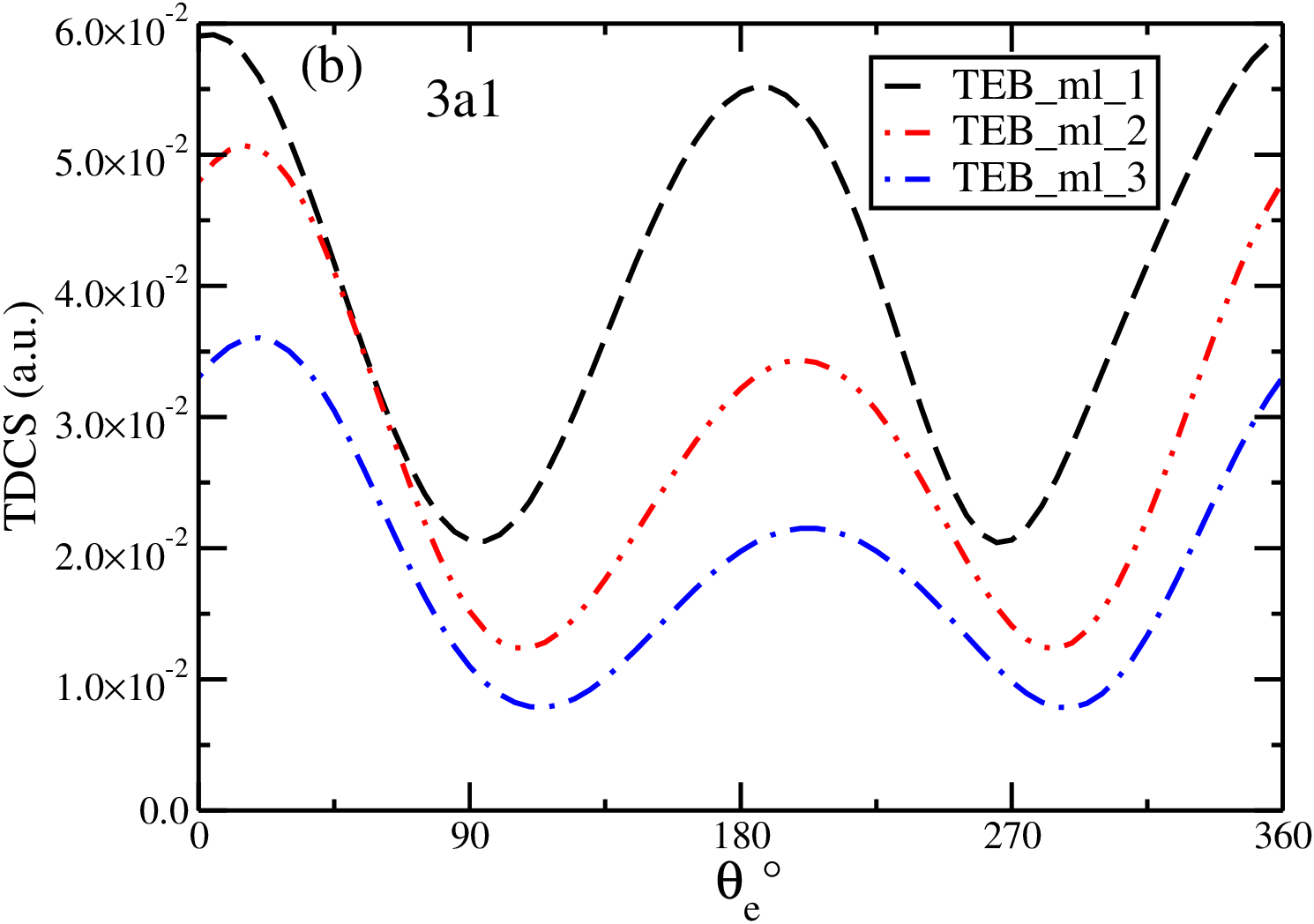}\ & \includegraphics[width=8.00cm]{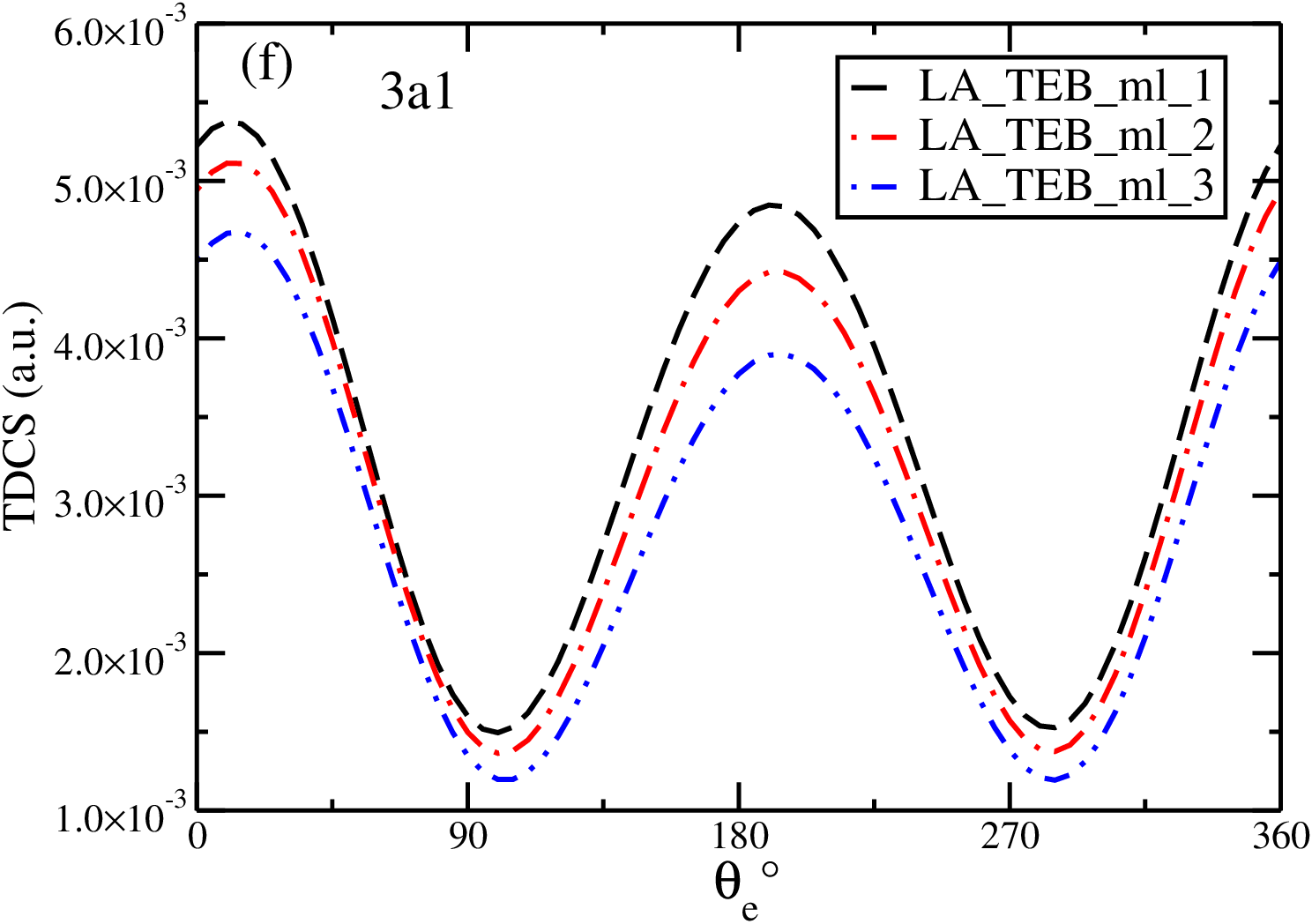}

				\end{tabular}
				
				\begin{tabular}{cc}
					\includegraphics[width=8.00cm]{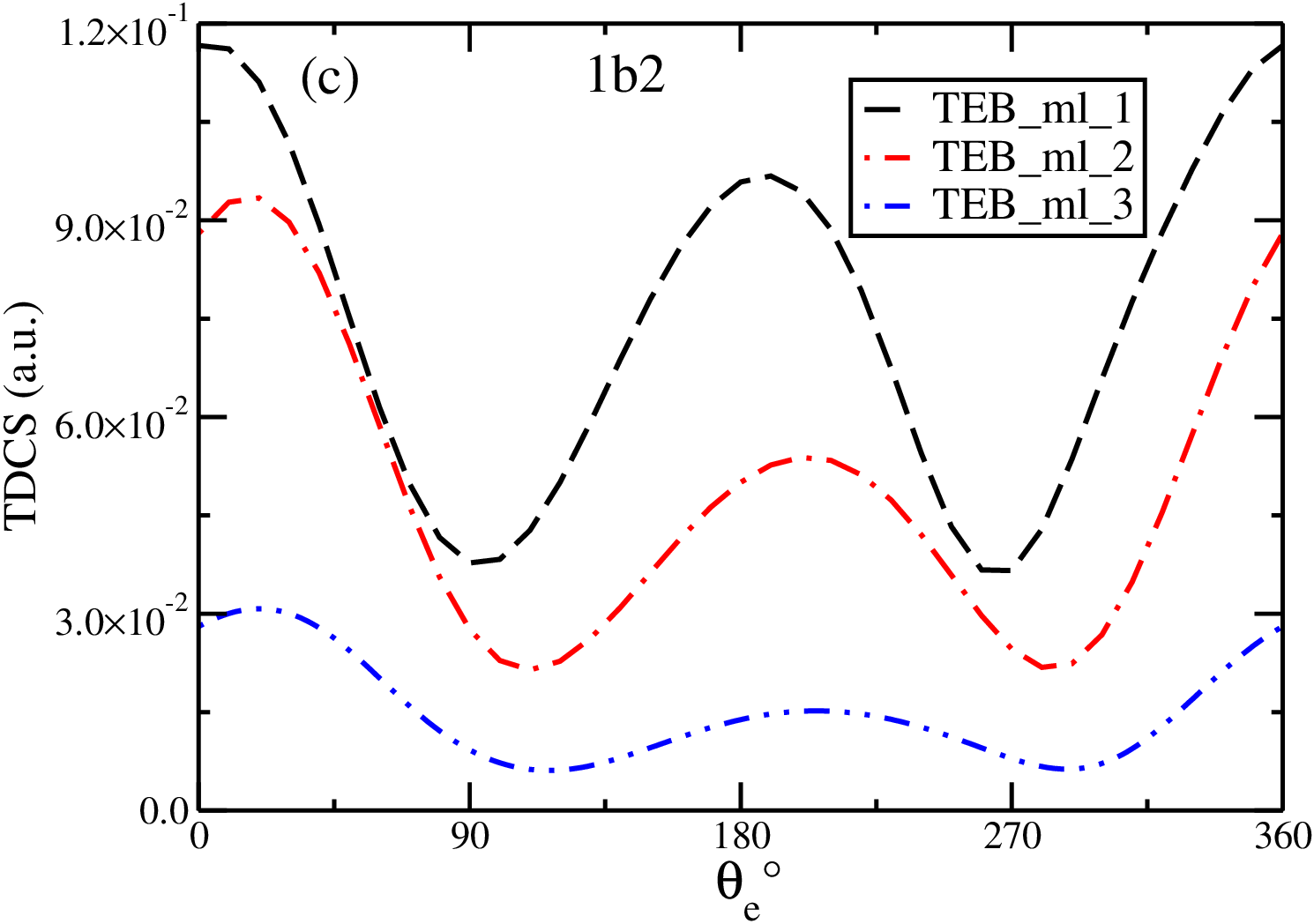}\ & \includegraphics[width=8.00cm]{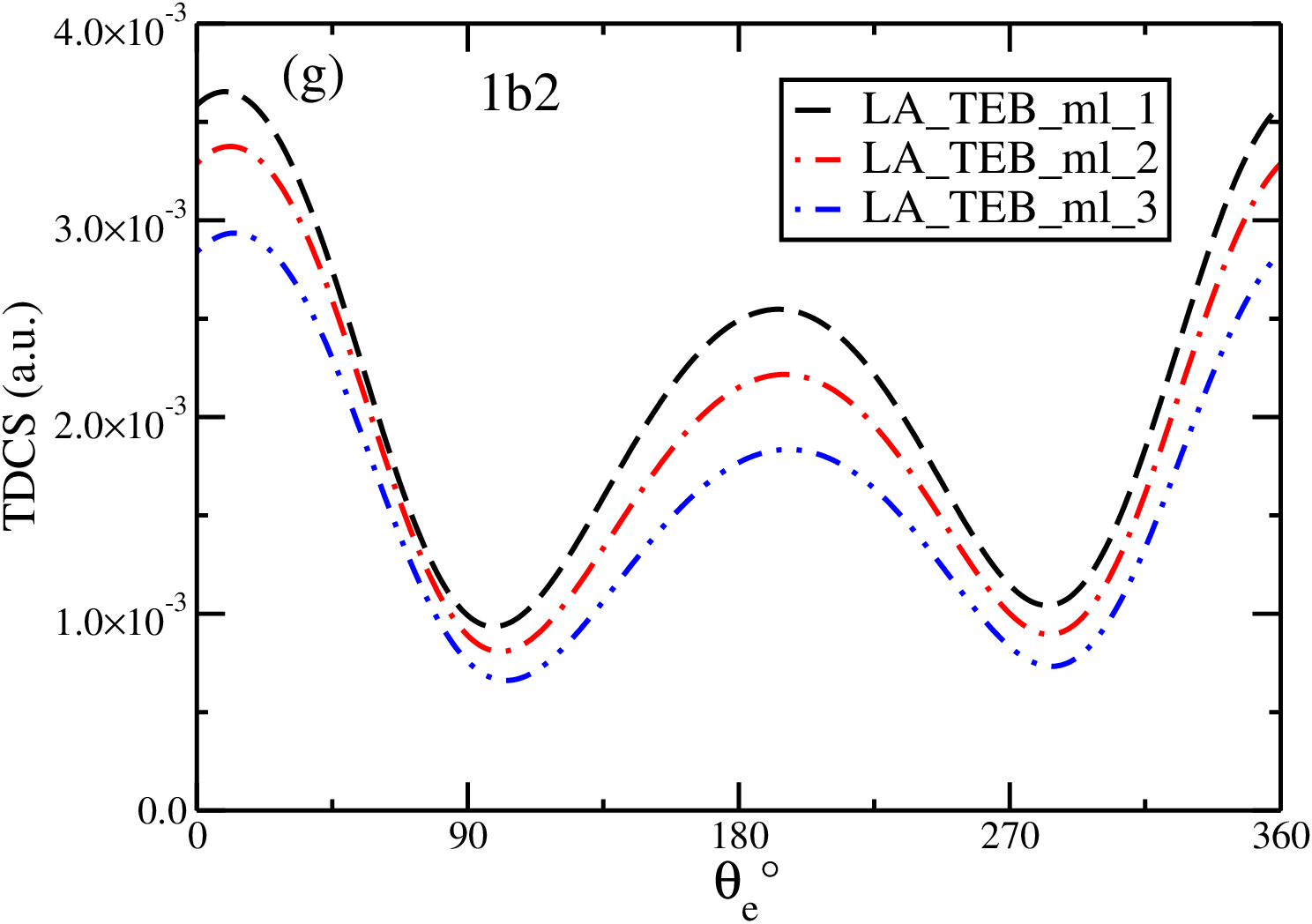}
					
				\end{tabular}
                \begin{tabular}{cc}
					\includegraphics[width=8.00cm]{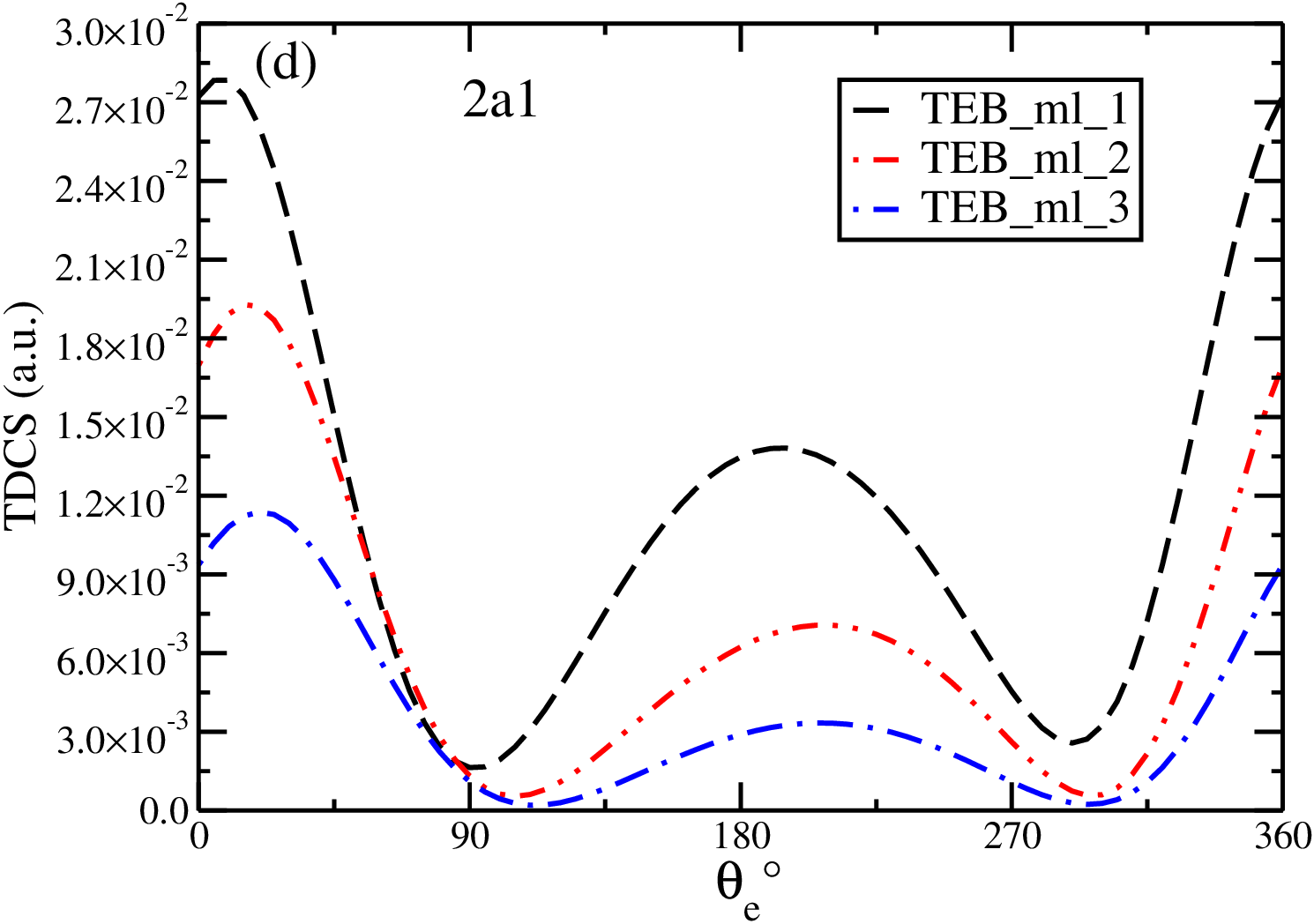}\ & \includegraphics[width=8.00cm]{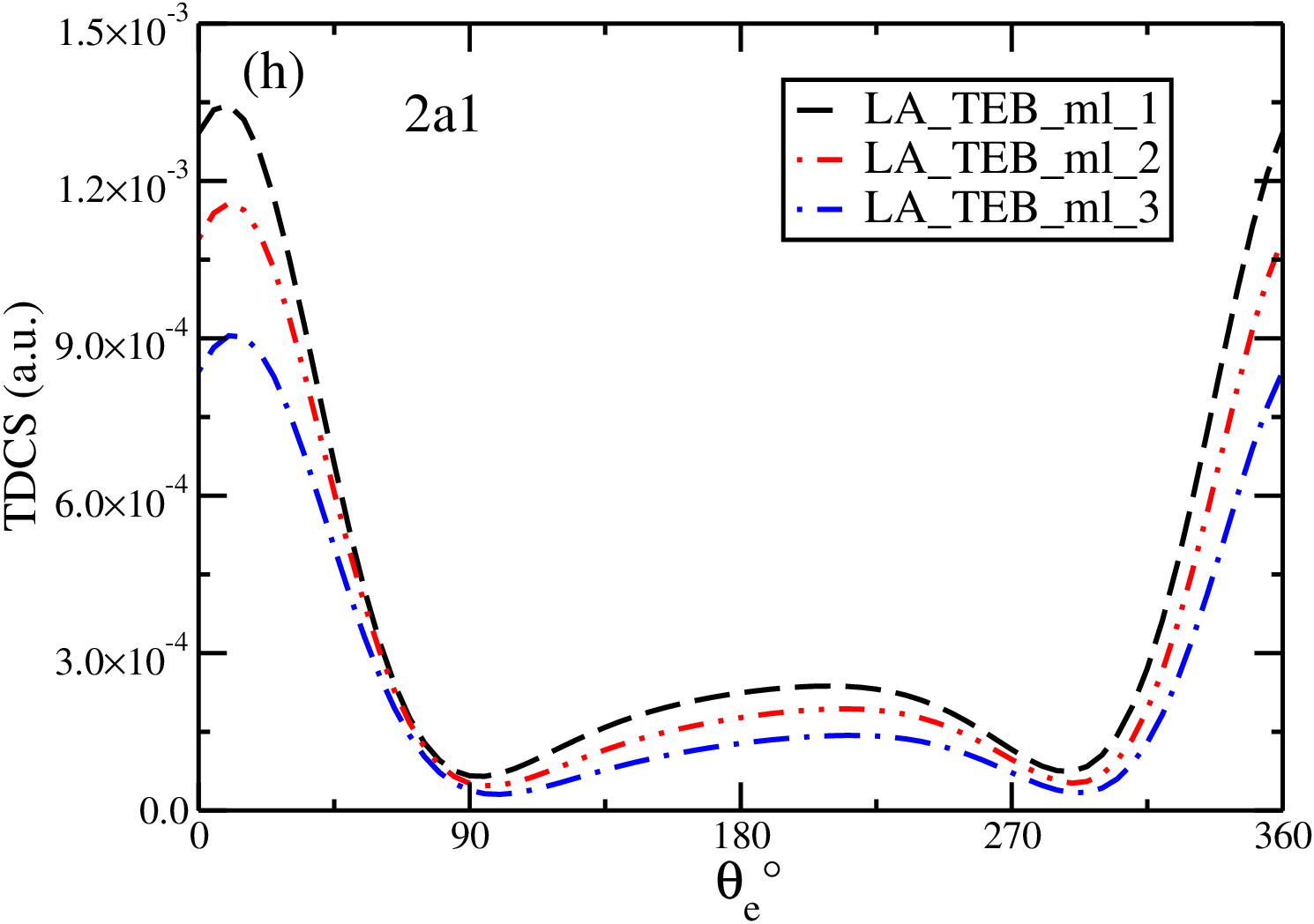}
				
				\end{tabular}

				\caption{TDCS as a function of $\theta_e$ for the (e,2e) process by TEB on $H_2O$ molecule. In the left column, Figs (a)-(d) represent the results of TEB without laser field for $1b_1$ in Fig (a), for $3a_1$ in Fig (b), for $1b_2$ in Fig (c), and $2a_1$ in Fig (d). Figs (e) - (h), the right column represents the result for the laser-assisted (e,2e) process with TEB, for $1b_1$ in Fig (e), for $3a_1$ in Fig (f), for $1b_2$ in Fig (g), and $2a_1$ in Fig (h). TDCS is calculated for OAM, $m_l$ = 1, $m_l$ = 2 and $m_l$ = 3 represented by a black dashed curve, red dashed-dotted-dotted curve, and blue dashed-dotted curve respectively at incident energy $E_i$ = 250eV, ejected electron energy $E_e$ 10 eV (except 8eV for $3a_1$) and $\theta_s$ = $\theta_p$ = 15$^\circ$.}   
			\label{fig:2}
			\end{figure*}
             We present the results of TDCS by the TEB without laser field for orbitals $1b_1$ in Fig \ref{fig:2}(a), $3a_1$ in Fig \ref{fig:2}(b), $1b_2$ in Fig$\ref{fig:2}$(c) and $2a_1$ in Fig \ref{fig:2}(d) (left panel of the fig 2). 
            We also present the TDCS for the laser-assisted (e,2e) process with twisted electron beam (LA-TEB) for the orbitals $1b_1$ in Fig \ref{fig:2}(e), $3a_1$ in Fig $\ref{fig:2}$(f), $1b_2$ in Fig \ref{fig:2}(g) and $2a_1$ in Fig \ref{fig:2}(h) (see right panel of the fig 2). 
           
            In the angular distribution of the TDCS by PW for the orbitals with the $p-like$ character, $1b_1$, $3a_1$, and $1b_2$, we observed two peaks in the binary region near $\theta_e$ = 90$^\circ$ (see the black solid curve in Fig \ref{fig:1}(a)-(c)) and a recoil peak near $\theta_e$ = 270$^\circ$ (see the black solid curve in Fig \ref{fig:1}(a)-(c)). On contrast of this, in the angular distribution of TDCS by TEB, for both the without (see Figs \ref{fig:2}(a)-(d)) and with laser-field results (see Figs \ref{fig:2}(e)-(h)), we observed two peak structure; a forward peak near $\theta_e$ = 0$^\circ$ (or 360$^\circ$) and a backward peak near $\theta_e$ = 180$^\circ$. Since the twisted electron beam is a superposition of plane waves, there is not a single well-defined momentum transfer from the twisted electron to the target. The momentum transfer vector contains both longitudinal and transverse components as can be seen from equation (\ref{27}) - (\ref{28}). This leads to different momentum transfer as the phase of the twisted electron changes. This results in destroying the symmetry about the plane wave momentum transfer resulting in the dip at $\theta_\Delta$. Additionally, the phase of the twisted electron beam is influenced by the OAM ($m_l$), varying accordingly for different $m_l$ values. As a result of the extra transverse component in the incident momentum vector and the dependence on the OAM number, the characteristic two-peak structure observed in the plane-wave case disappears for twisted electrons \cite{harris2019ionization}.
           
            From Figs \ref{fig:2}(a)-(d) and \ref{fig:2}(e)-(h), we observed that the magnitude of TDCS reduces by order of one for laser-assisted ionization processes (see Figs \ref{fig:2}(a)-(d) and \ref{fig:2}(e)-(h)). This magnitude further reduces with the increase in the OAM from $m_l$ = 1 to $m_l$ = 3 (see blue dashed-dotted curves in Fig \ref{fig:2}(a)-(d) and Fig \ref{fig:2}(e)-(f)). We observed that the difference in TDCS magnitudes for $m_l$ = 1, $m_l$ = 2 and $m_l$ = 3 in the absence of a laser field (see the black dashed curve, red dash-dot-dot curve, and blue dash-dotted curve respectively in Figs. \ref{fig:2}(a)–(d)) is larger than the corresponding magnitude difference in the laser-assisted case (see the black dashed curve, red dashed-dotted-dotted curve, and blue dashed-dotted curve in Fig \ref{fig:2}(e)-(h)). For without laser-field study, the forward peak (near $\theta_e$ = 180$^\circ$) slightly shifts towards a larger ejected electron angle ($\theta_e$) when the OAM is increased (see the black dashed curve, red dash-dot-dot curve, and blue dash-dotted curve, respectively in Figs. \ref{fig:2}(a)–(d)). While no such shifts are observed for laser-assisted study (see Fig \ref{fig:2}(e)-(h)). We observed comparatively a symmetry about $\theta_s$ = $\theta_p$ = 15$^\circ$ in the angular distribution of TDCS for the laser-assisted case (see Fig \ref{fig:2}(e)-(h)). For the three orbitals with dominant $p$ character we observe a prominent contribution in the forward direction (see peaks around $\theta_e$ = 0$^\circ$ or 360$^\circ$ for dashed, dashed-dotted-dotted and dashed-dotted curves in Fig \ref{fig:2}) and also in the backward direction (see around $\theta_e$ = 180$^\circ$ in Fig \ref{fig:2}) for $m_l$ = 1,2 and 3. 
            For orbital $2a_1$ with dominant $s$ character
            the forward peaks get enhanced for $m_l$ = 1,2 and 3 (see peaks around $\theta_e$ = 0$^\circ$ or 360$^\circ$ for dashed, dashed-dotted-dotted and dashed-dotted curves in Fig \ref{fig:2} (d) and (h)) and the backward peaks get suppressed (see peak near $\theta_e$ = 180$^\circ$ in Fig \ref{fig:2}(d) and (h)).
           \subsection{Angular distribution of $(TDCS)_{av}$}

\begin{figure*}[htp!]
		\begin{tabular}{ccc}
			\includegraphics[width=6.00cm]{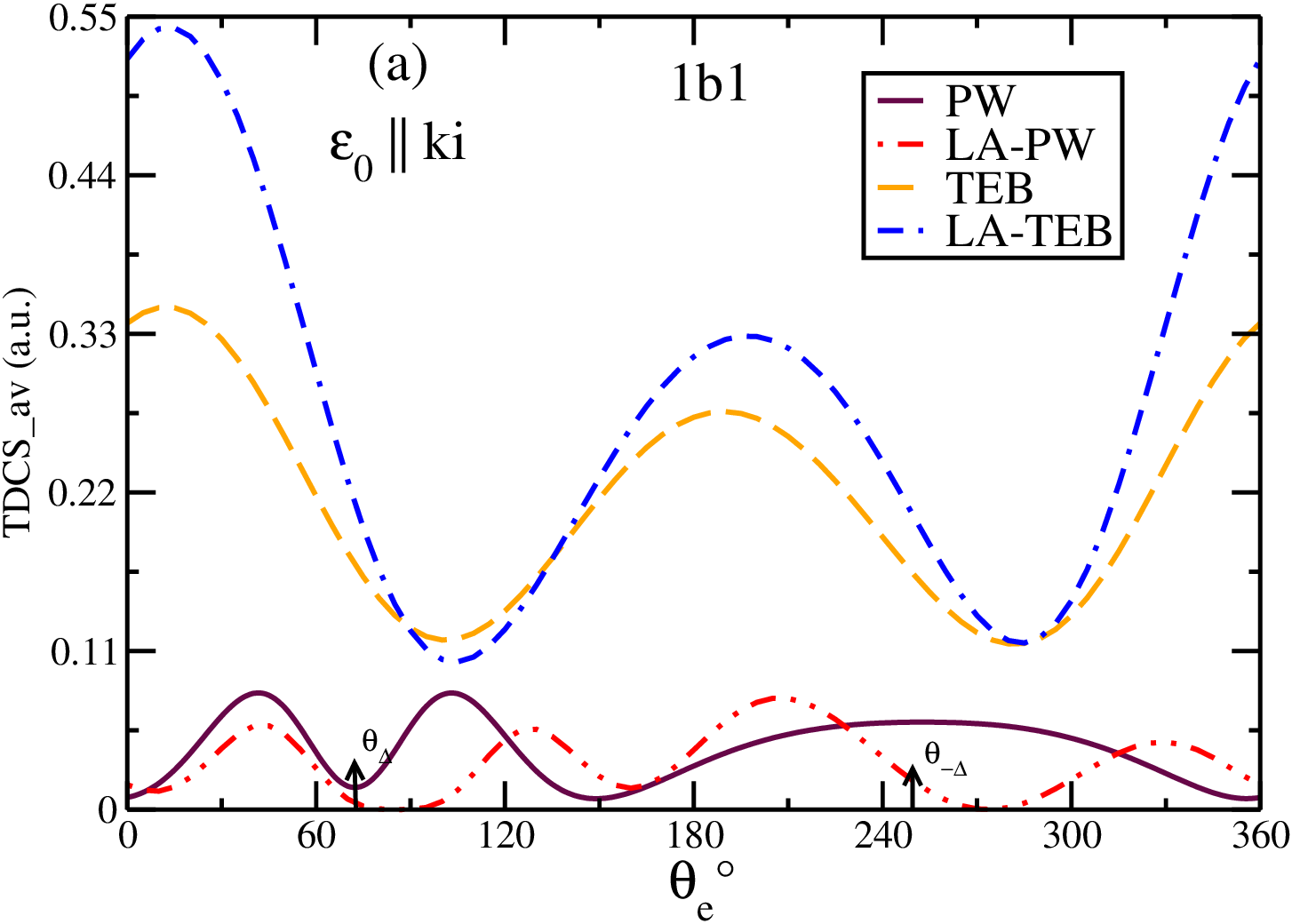}\ & \includegraphics[width=6.00cm]{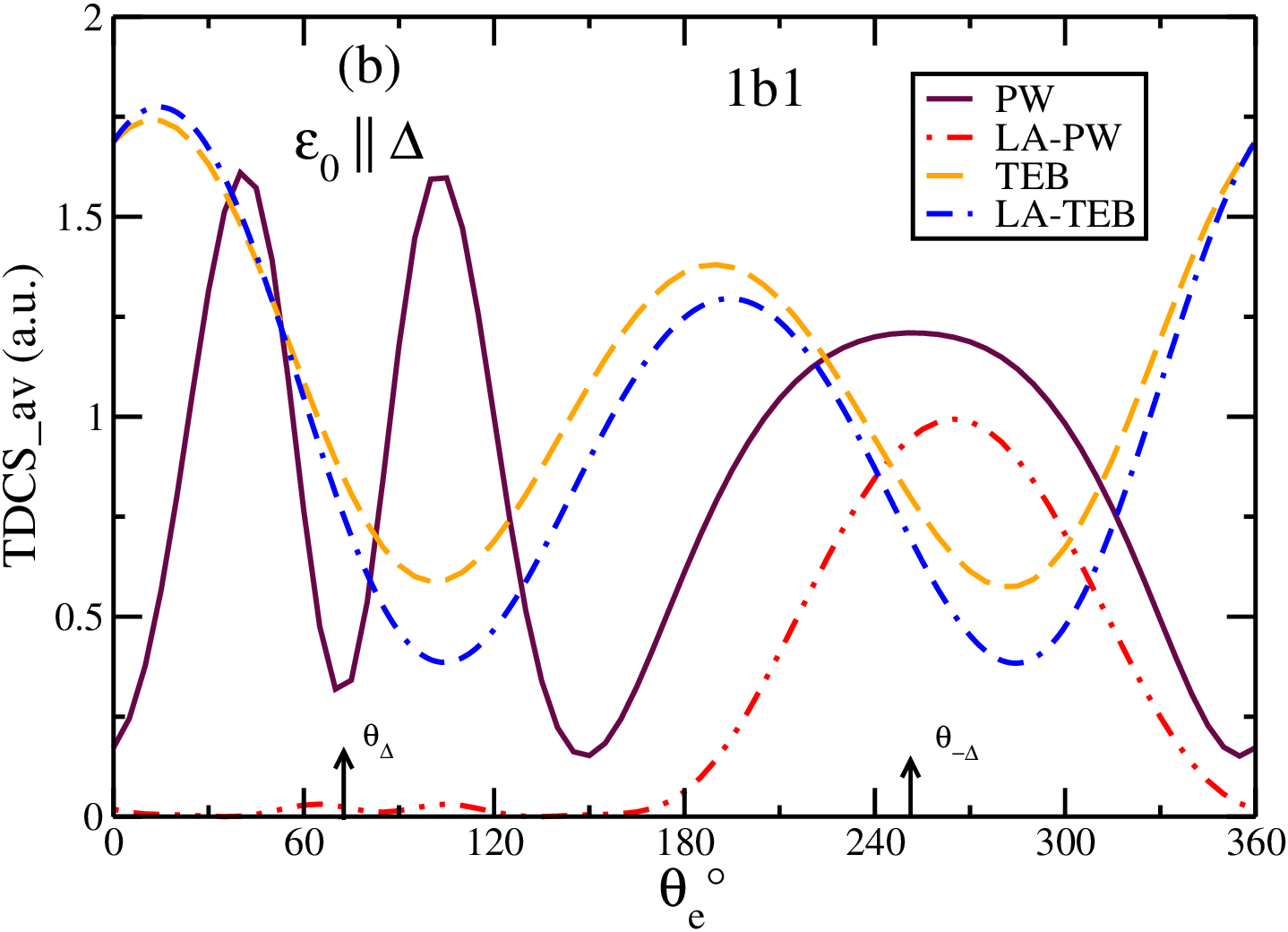}\
			& \includegraphics[width=6.00cm]{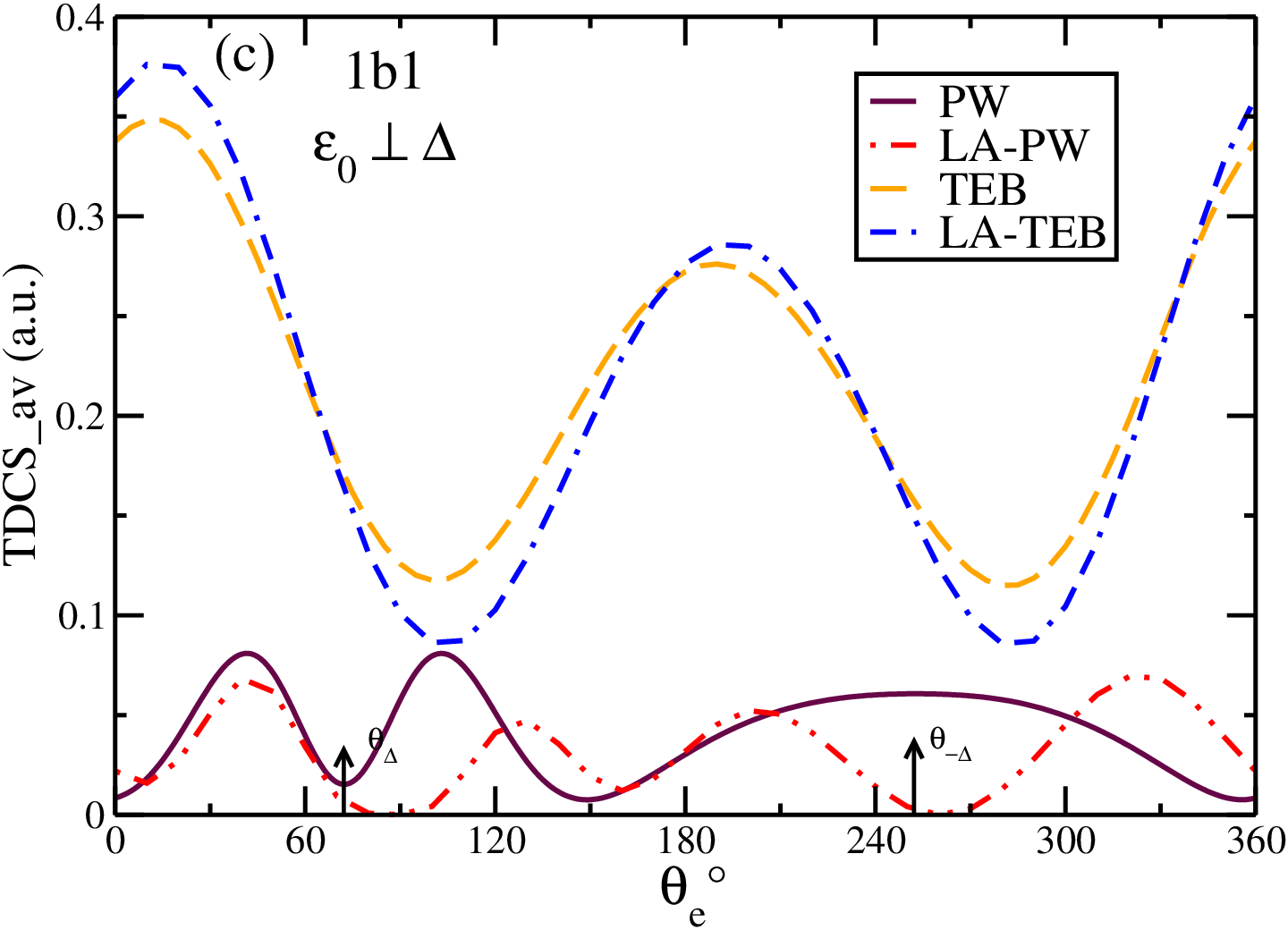}
	\end{tabular}
		\caption{$(TDCS)_{av}$ as a function of $\theta_e$ for the twisted electron (e, 2e) process on the $H_{2}O$ molecular target (sub-Fig (a) for outer orbital $1b_1$ with $\varepsilon_0$ $\parallel$ $k_i$, sub-Fig (b) for outer orbital $1b_1$ with $\varepsilon_0$ $\parallel$ $\Delta$ , and sub Fig (c) for outer orbital $1b_1$ with $\varepsilon_0$ $\perp$ $\Delta$).The kinematics is the same as in Figure 1. Keeping $\theta_{s}$ = $\theta_{p}$ = 15$^\circ$. The solid maroon curve for PW without laser-field, red dashed-dotted-dotted curve for LA-PW, orange dashed curve for TEB , and blue dashed-dashed-dotted curve for LA-TEB. In Fig. 3(a), the TDCS for LA-PW is multiplied by 100, while LA-TEB is multiplied by 8. In Fig. 3(b), PW  is multiplied by 20, LA-PW by 300, and TEB by 5. In Fig. 3(c), the multiplicative factors are 100 for LA-PW and 6 for LA-TEB.}
		 \label{fig:3}
	\end{figure*}

\begin{figure*}[htp!]
		\begin{tabular}{ccc}
			\includegraphics[width=6.00cm]{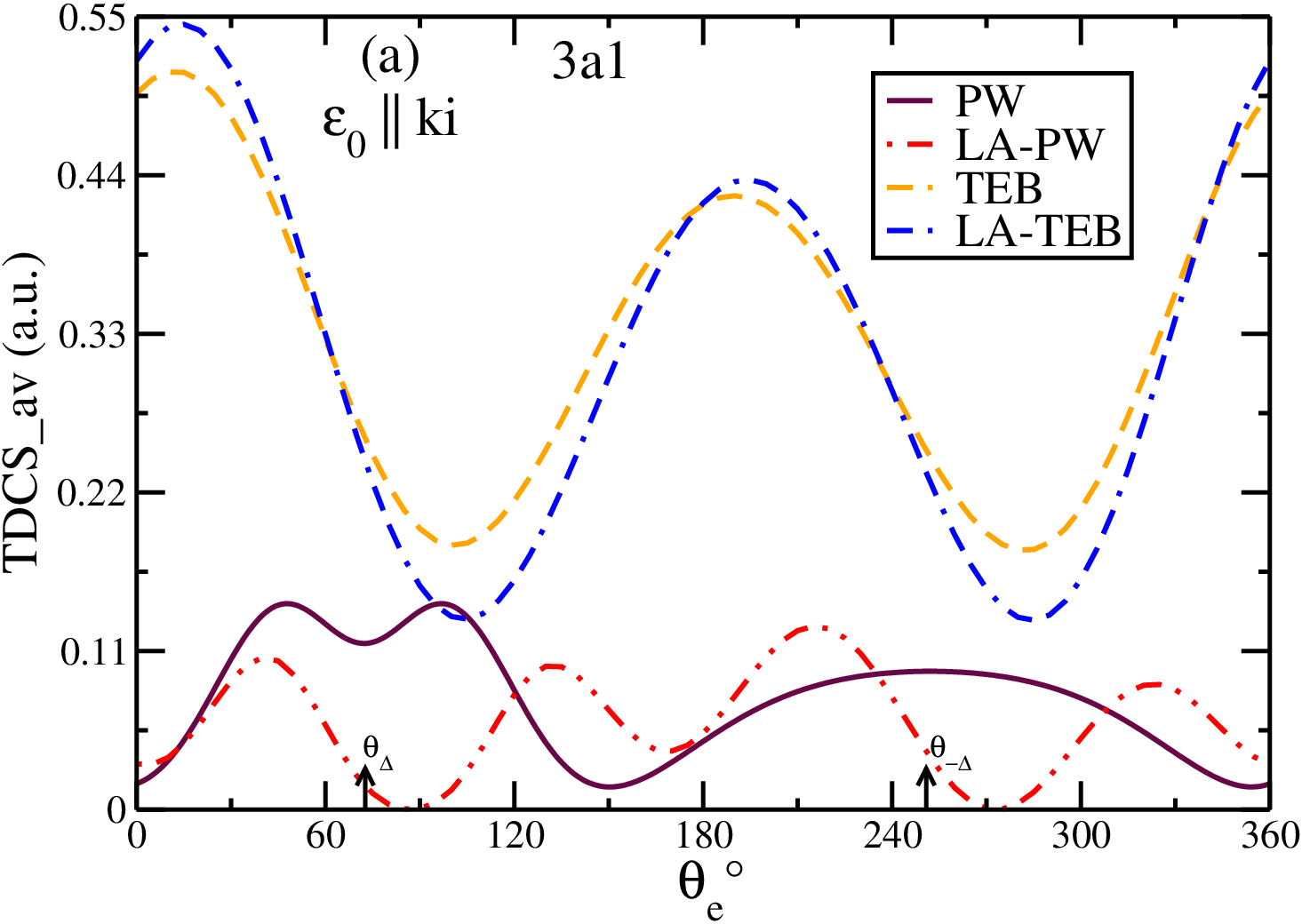}\ & \includegraphics[width=6.00cm]{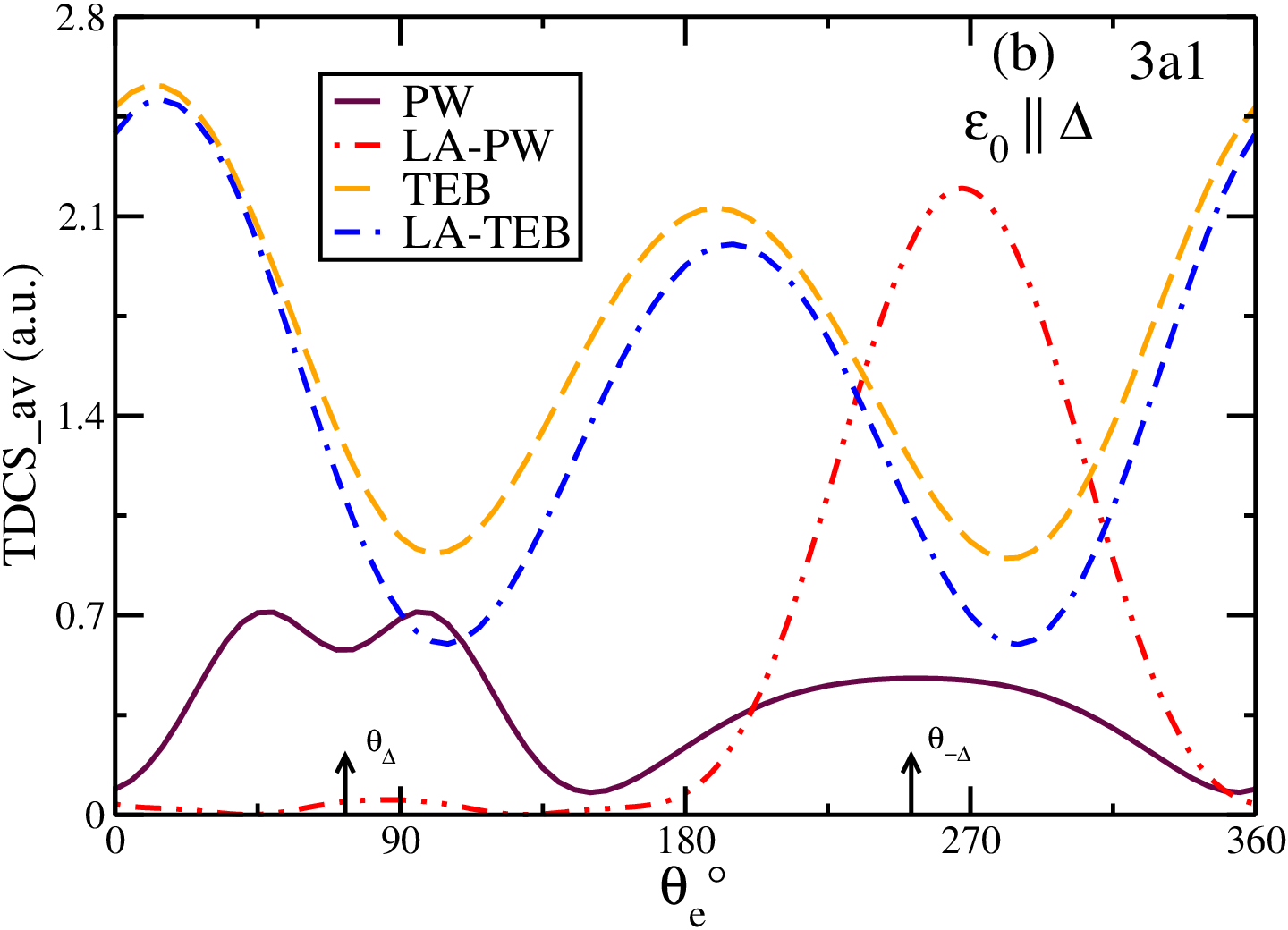}\
			& \includegraphics[width=6.00cm]{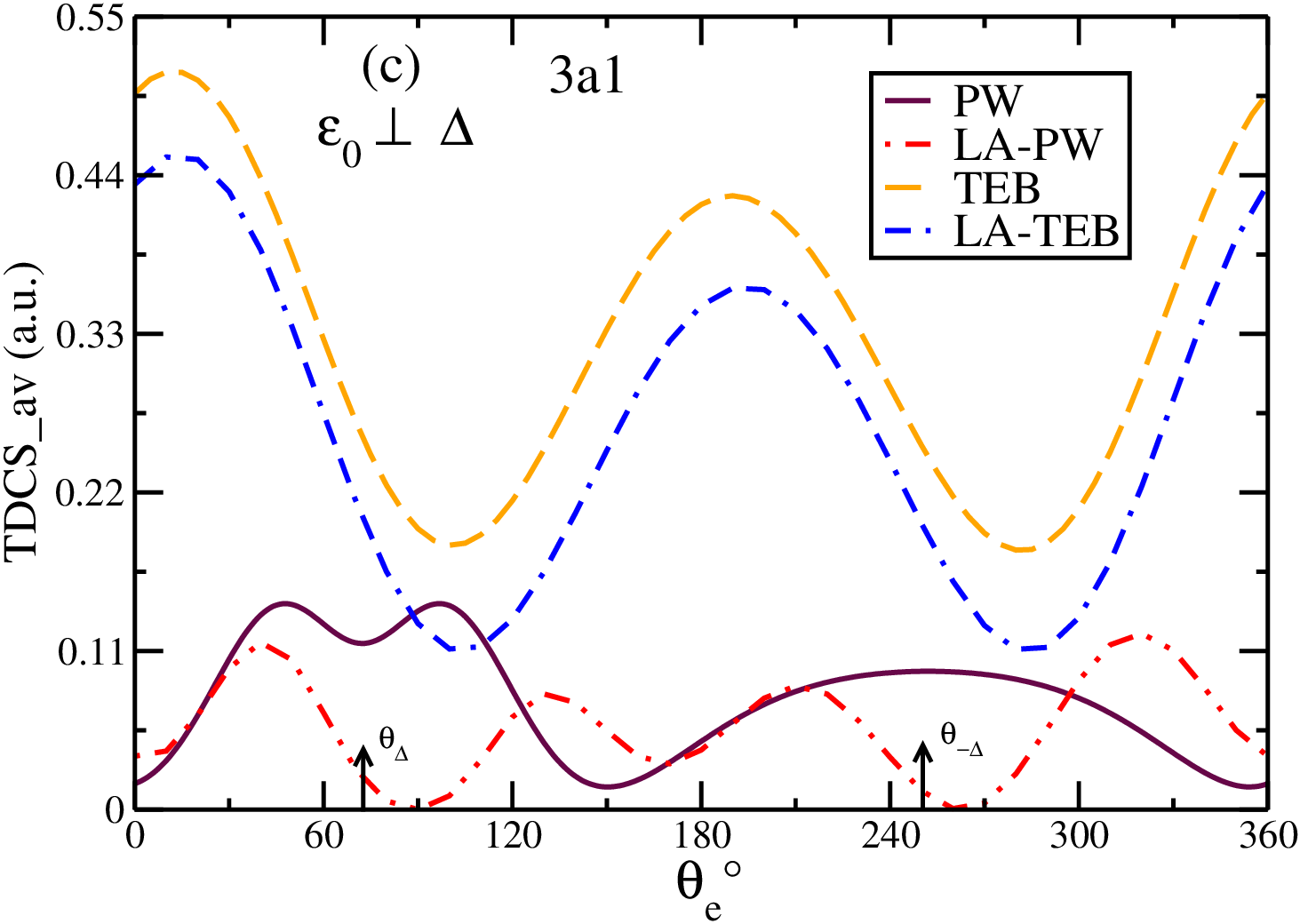}
	\end{tabular}
		\caption{Same as Fig \ref{fig:3} except for the outer orbital $3a_1$. In Fig. 3(a), the TDCS for LA-PW is multiplied by 100, while LA-TEB is multiplied by 6. Fig. 3(b) shows LA-PW by 250 and TEB by 5. In Fig. 4(c), the multiplicative factors are 100 for LA-PW and 5 for LA-TEB.}
		 \label{fig:4}
	\end{figure*}

\begin{figure*}[htp!]
		\begin{tabular}{ccc}
			\includegraphics[width=6.00cm]{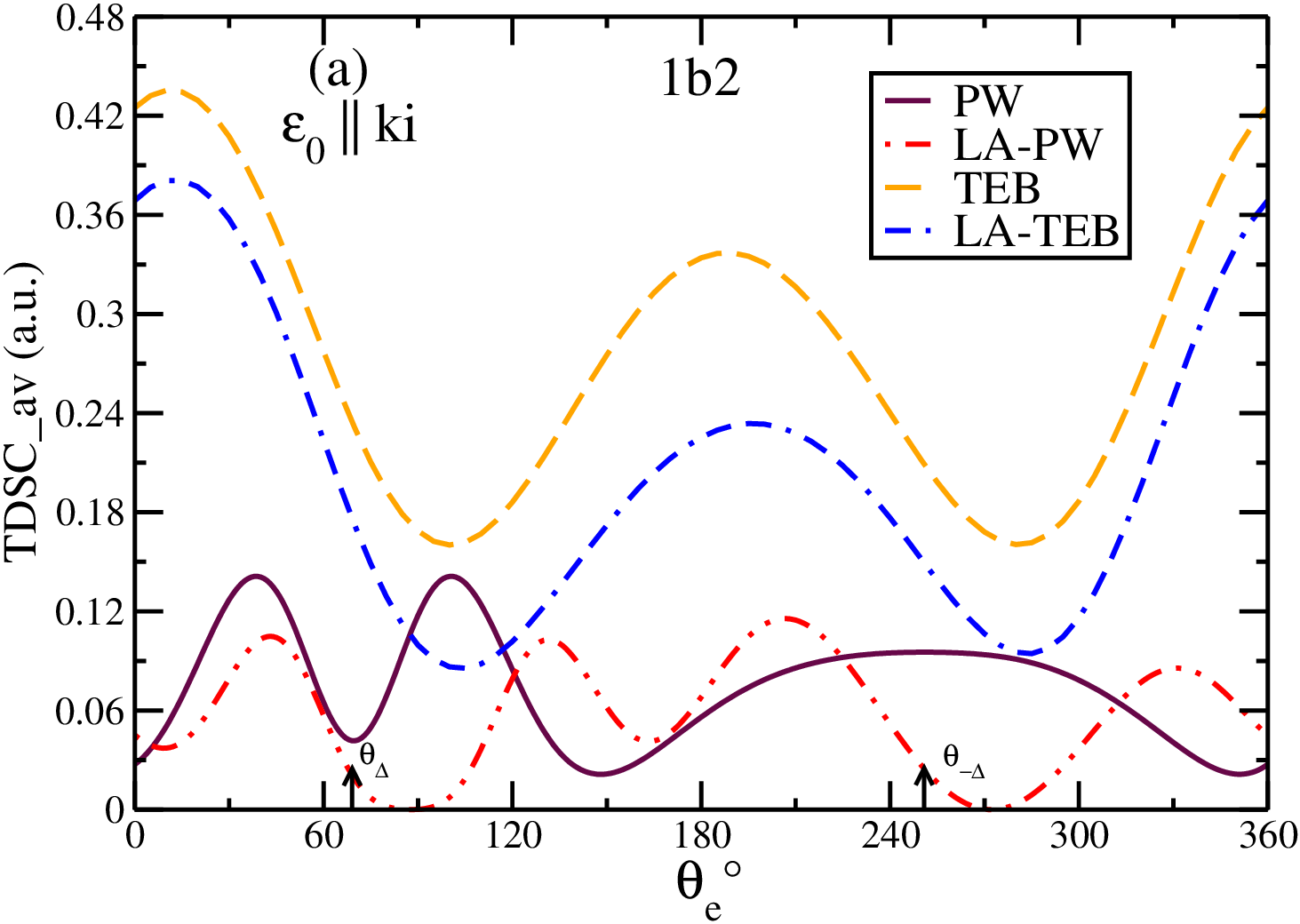}\ & \includegraphics[width=6.00cm]{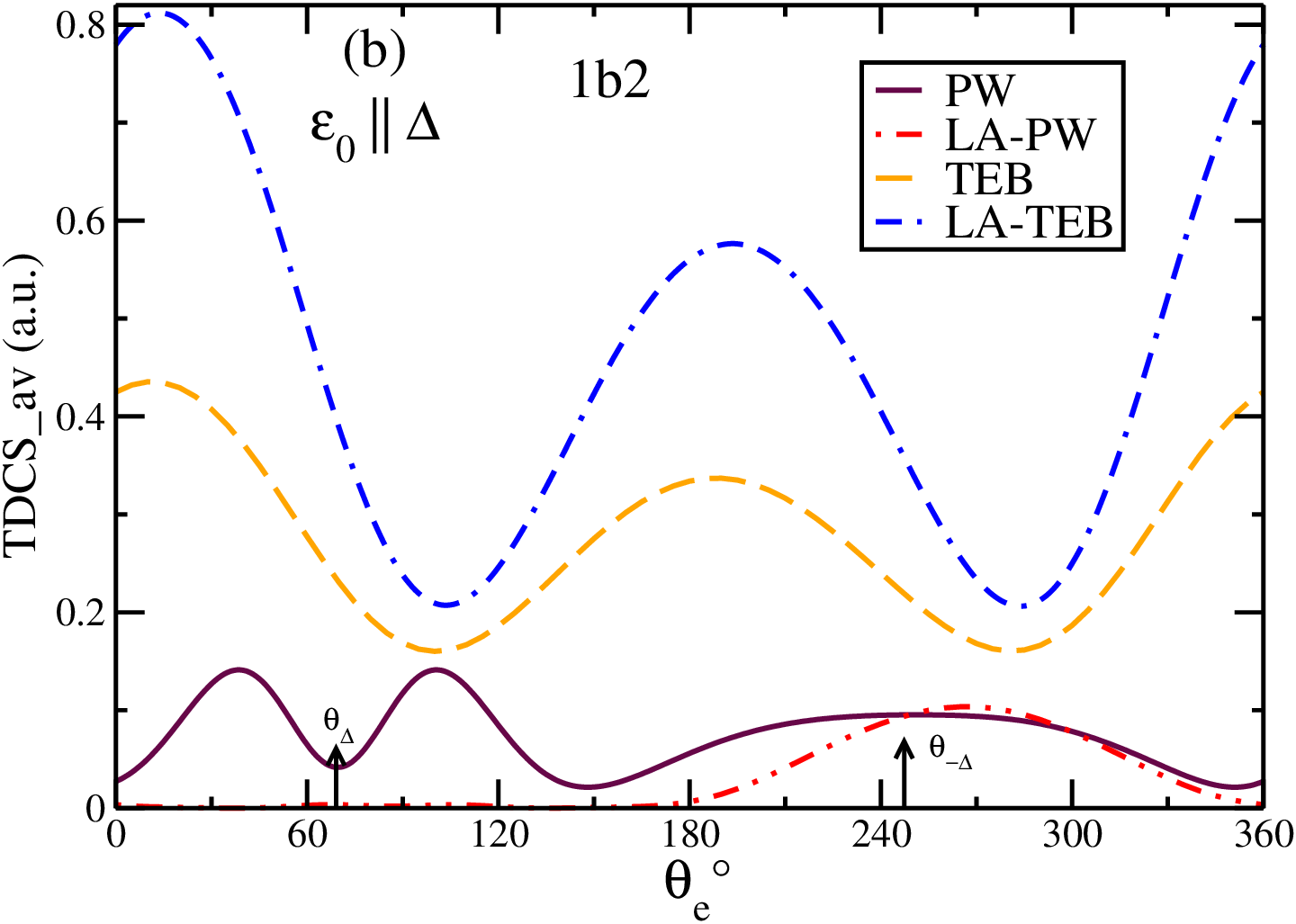}\
			& \includegraphics[width=6.00cm]{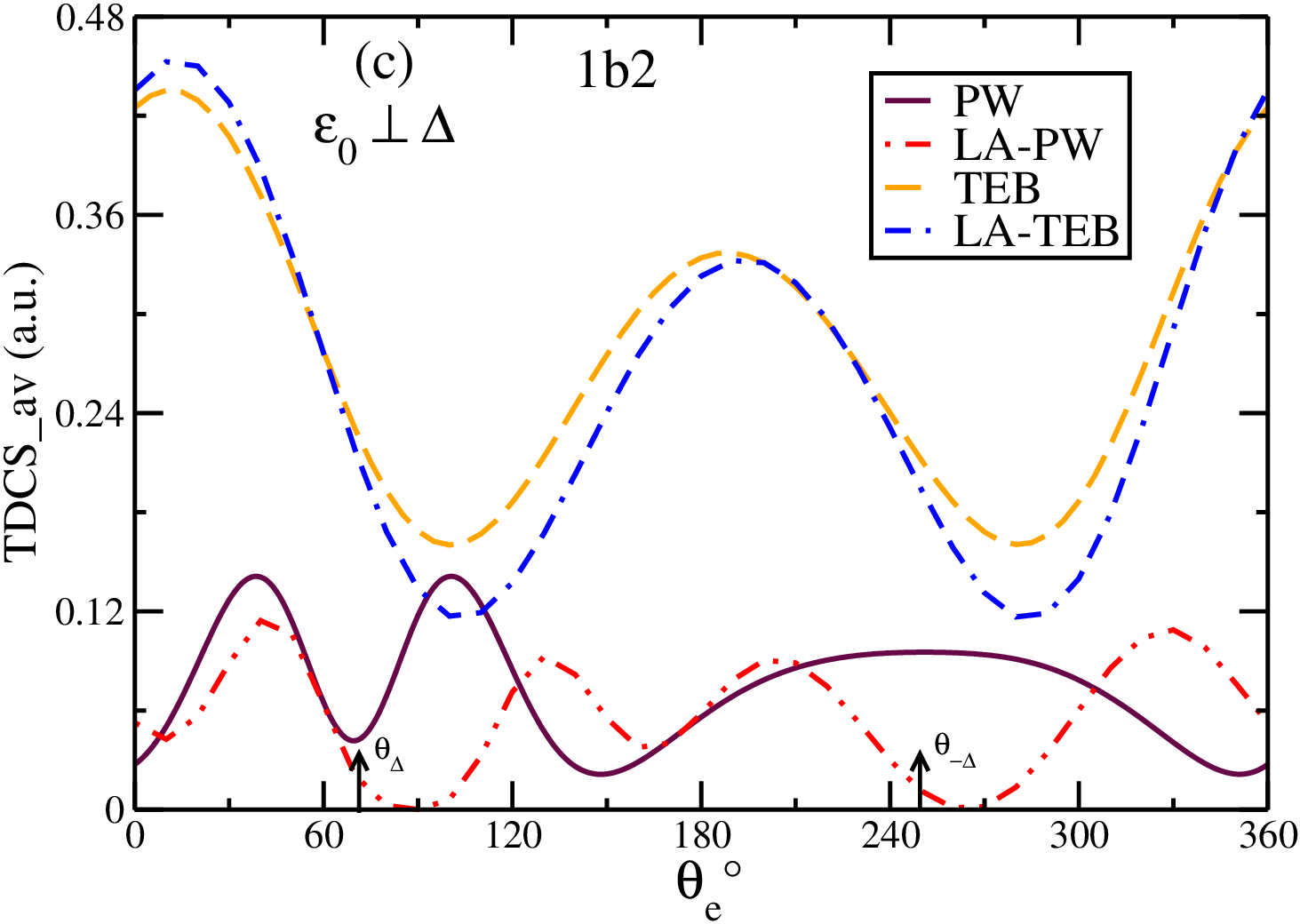}
	\end{tabular}
		\caption{Same as Fig \ref{fig:3} except for the outer orbital $1b_2$. In Fig. 5(a), the TDCS for LA-PW is multiplied by 100, while LA-TEB is multiplied by 8. Fig. 5(b) shows LA-PW by 100 and LA-TEB by 8. In Fig. 5(c), the multiplicative factors are 100 for LA-PW and 6 for LA-TEB.}
		 \label{fig:5}
	\end{figure*}

\begin{figure*}[htp!]
		\begin{tabular}{ccc}
			\includegraphics[width=6.00cm]{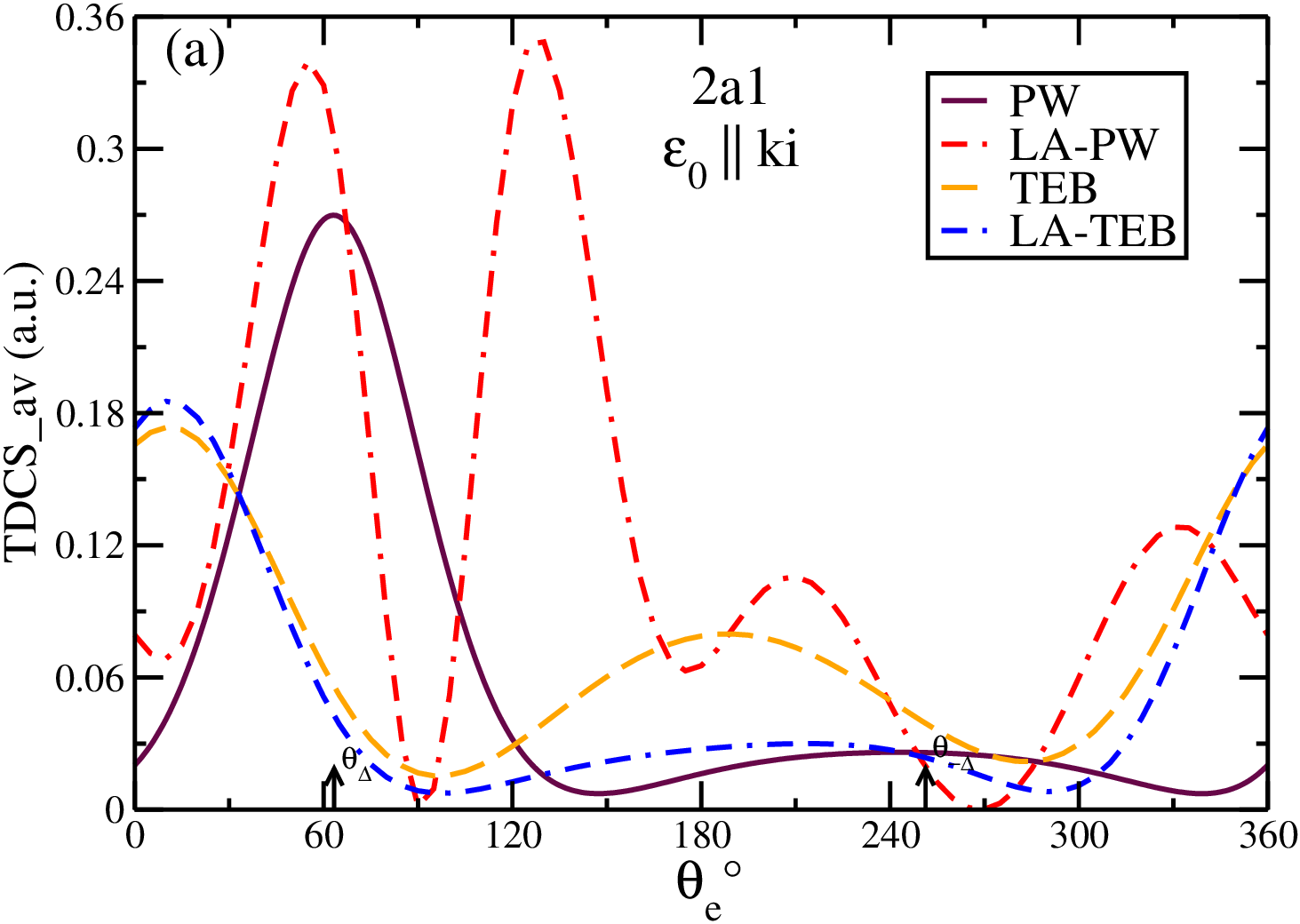}\ & \includegraphics[width=6.00cm]{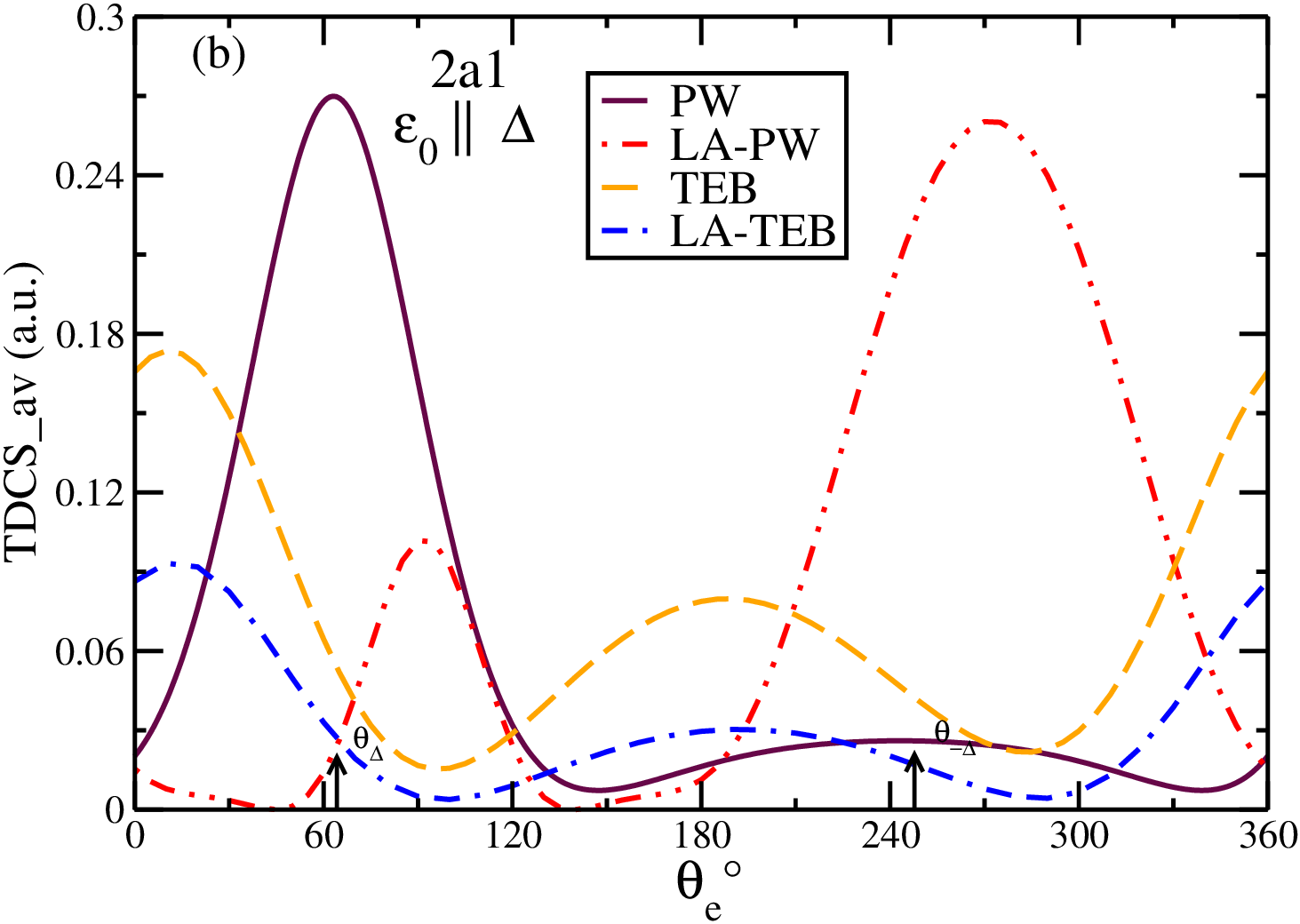}\
			& \includegraphics[width=6.00cm]{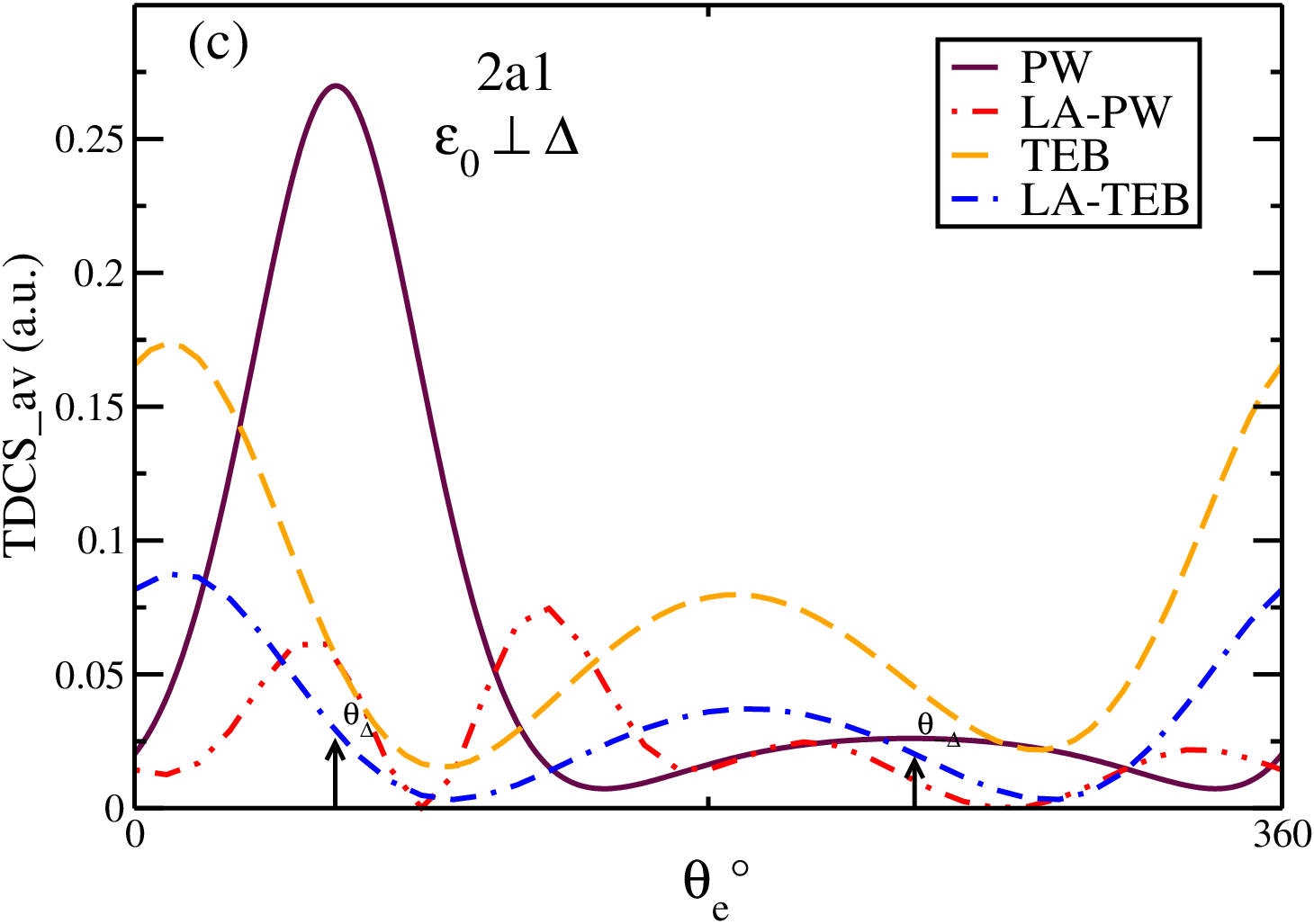}
	\end{tabular}
		\caption{Same as Fig \ref{fig:3} except for the outer orbital $2a_1$. In Fig. 6(a), the TDCS for LA-PW is multiplied by 500, while LA-TEB is multiplied by 15. In Fig. 6(b), LA-PW by 300. In Fig. 6(c), the multiplicative factors are 100 for LA-PW and 10 for LA-TEB.}
		 \label{fig:6}
	\end{figure*}

        In Figs \ref{fig:3} - \ref{fig:6}, we present the $(TDCS)_{av}$ as a function of ejected electron angle $\theta_e$. The calculations are shown for the orbital $1b_1$ in Fig \ref{fig:3}, $3a_1$ in Fig \ref{fig:4}, $1b_2$ in Fig \ref{fig:5} and $2a_1$ in Fig \ref{fig:6}. The kinematics used here is $E_{i}$ = 250eV, $E_{e}$ = 10eV (8eV for $3a_1$), $\theta_s$ = $\theta_p$ = 15$^\circ$ for the three laser field orientations; namely $\varepsilon_0$ $\parallel$ $k_i$, $\varepsilon_0$ $\parallel$ $\Delta$ and $\varepsilon_0$ $\perp$ $\Delta$. (a), (b) and (c) frames of each figure represent the results of $(TDCS)_{av}$ for laser field orientations $\varepsilon_0$ $\parallel$ $k_i$, $\varepsilon_0$ $\parallel$ $\Delta$ and $\varepsilon_0$ $\perp$ $\Delta$ respectively. In Figs \ref{fig:3} - \ref{fig:6}, we compare the angular profile of $(TDCS)_{av}$ for PW (maroon solid curve), LA-PW (red dashed-dotted-dotted curve), TEB (orange dashed curve), and LA-TEB (blue dashed-dashed-dotted curve). 
           We observed that the magnitude of the $(TDCS)_{av}$ was enhanced for the TEB and LA-TEB compared to that for the PW (see orange dashed and blue dashed-dotted-dotted curves in Figs \ref{fig:3} - \ref{fig:5}). In Figs \ref{fig:3} - \ref{fig:5} (subsequent figures (a), (b), and (c)) for $p$ character orbitals, the dual-peaks observed for PW disappear in the angular profile of $(TDCS)_{av}$ for both LA-TEB and TEB calculations. In both cases, we observe prominently two peaks directed in the forward direction (near $\theta_e$ = 0$^\circ$ or 360$^\circ$) and backward direction ($\theta_e$ = 180$^\circ$). But for the orbital $2a_1$ with $s-like$ character, the magnitude of $(TDCS)_{av}$ for TEB is less than that of the PW (see Fig. \ref{fig:6}). For $2a_1$ orbital LA-TEB, the backward peak was suppressed compared to the TEB results (see blue dashed-dotted-dotted and orange dashed curves in Fig \ref{fig:6}.)

    \begin{figure*}[htp!]

   \subsection{Impact Parameter effects on angular distribution of TDCS}

	       \begin{tabular}{cc}
					\includegraphics[width=8.00cm]{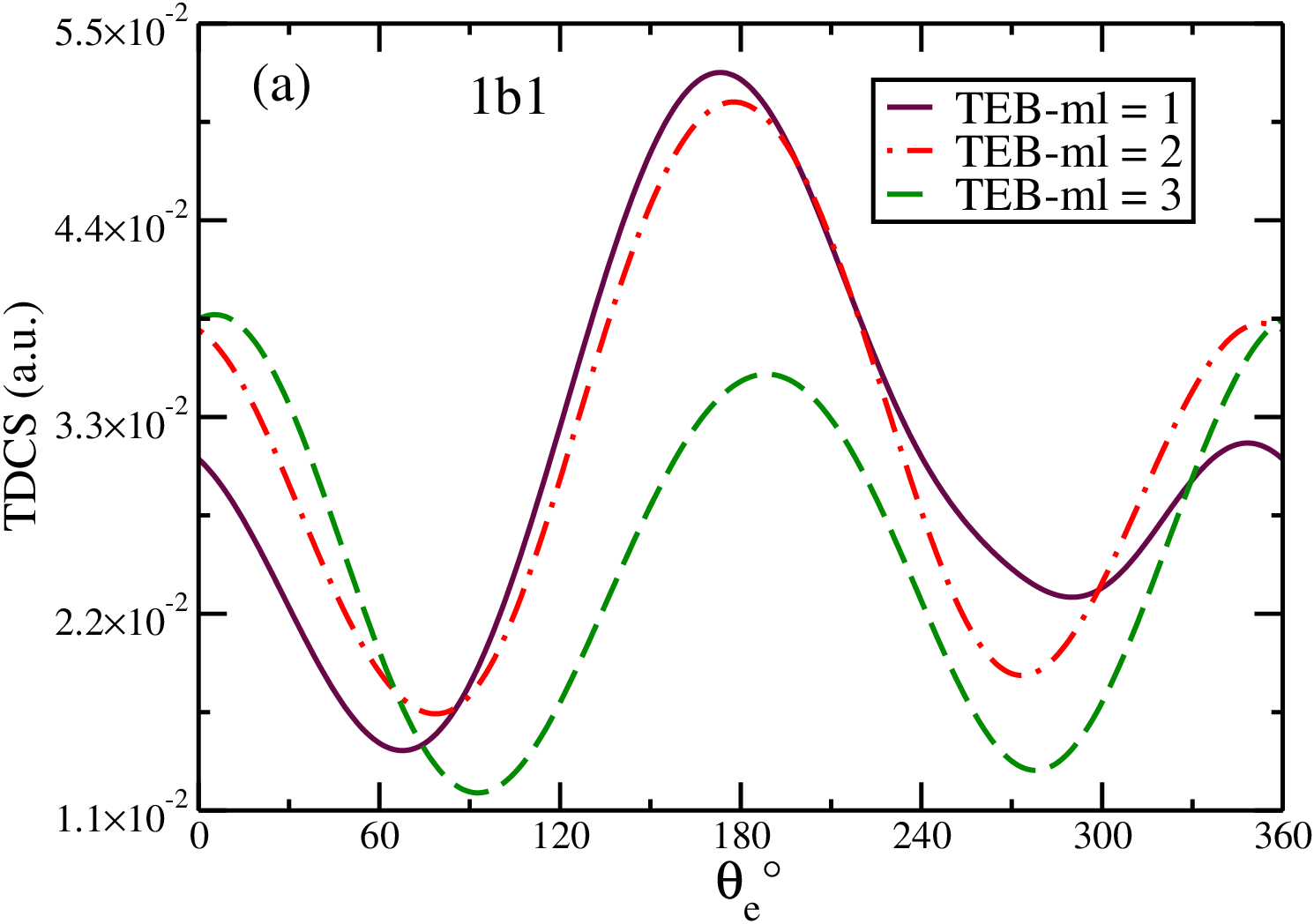}\ & \includegraphics[width=8.00cm]{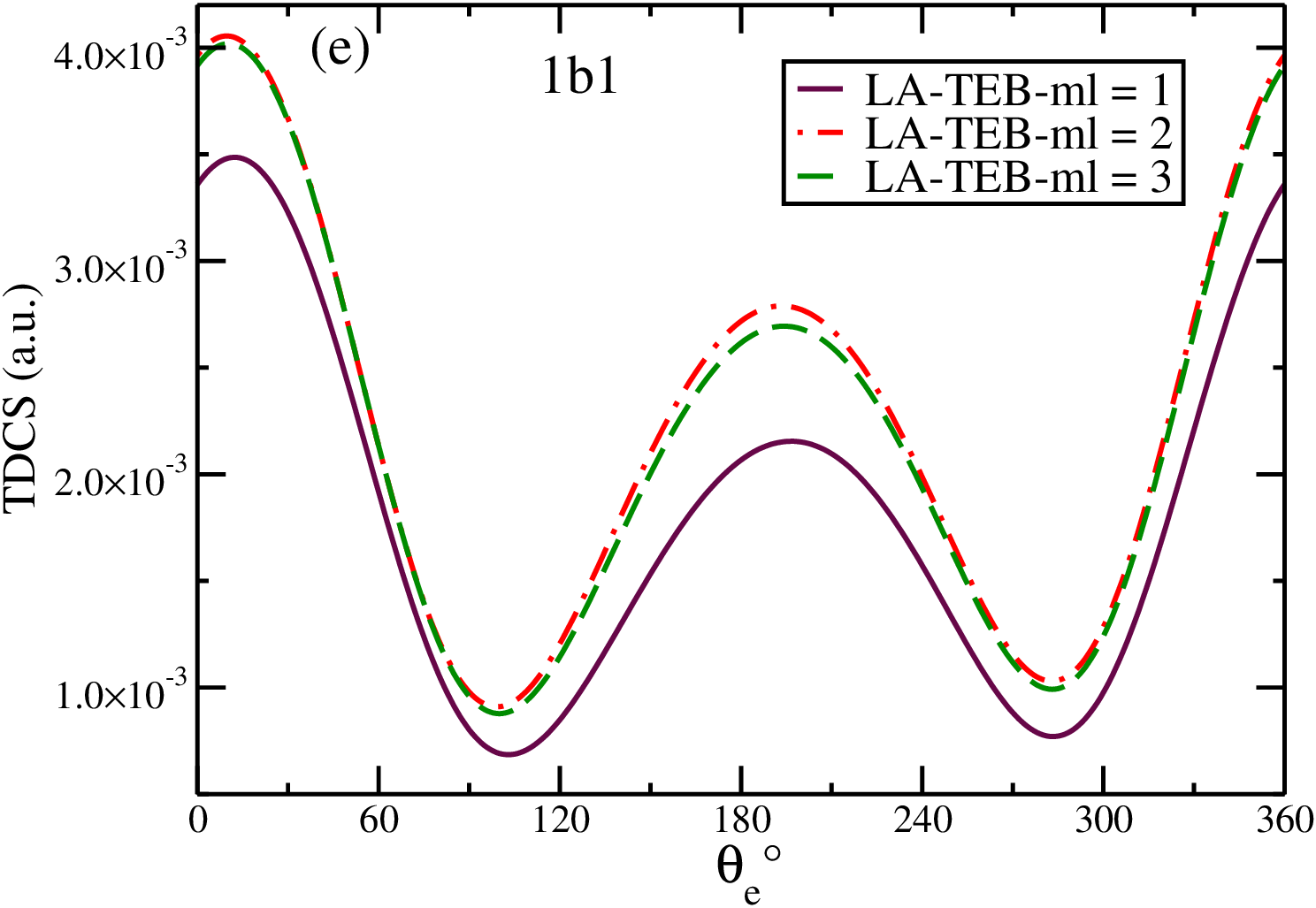}
					
					\end{tabular}
				
				\begin{tabular}{cc}
					\includegraphics[width=8.00cm]{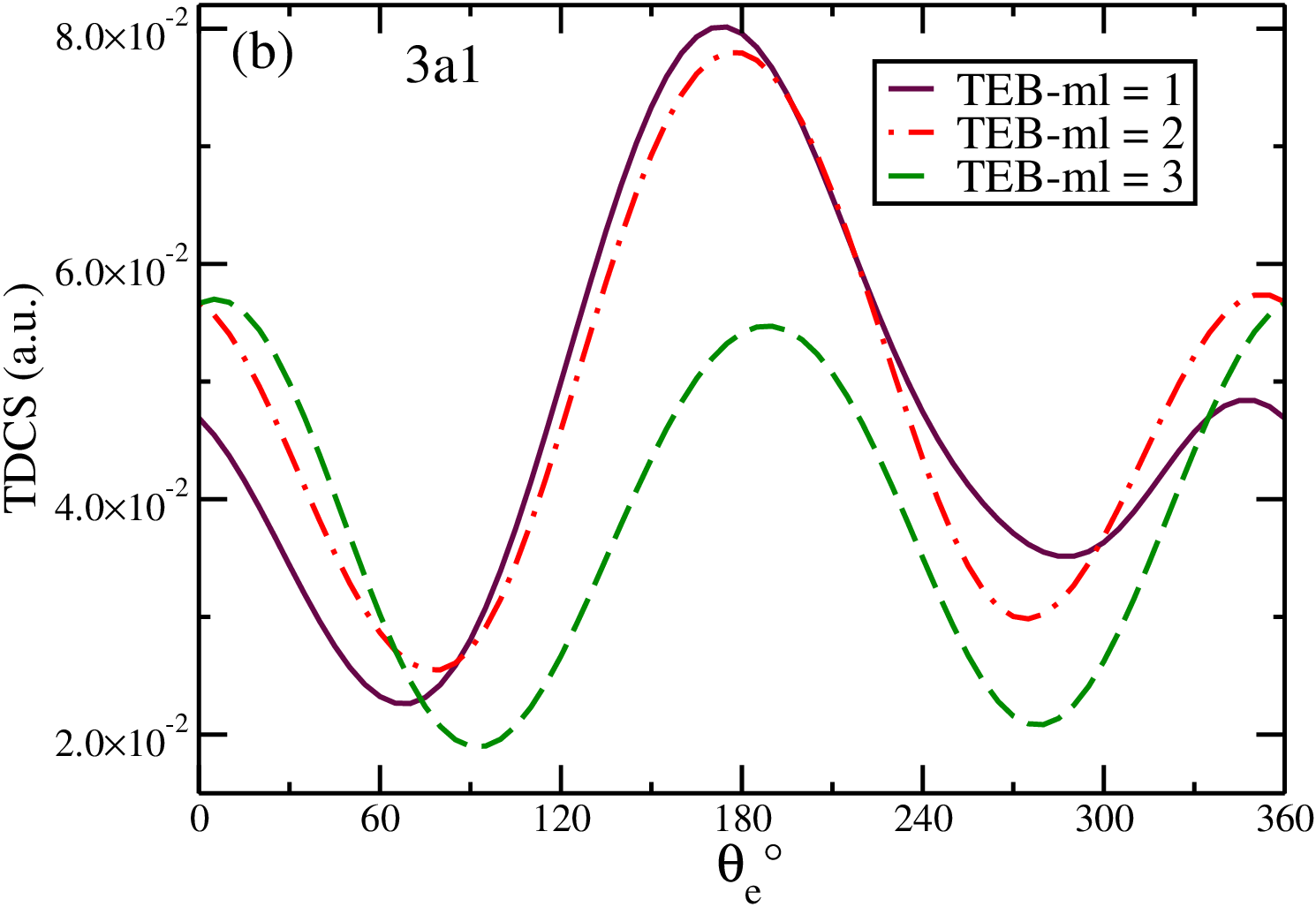}\ & \includegraphics[width=8.00cm]{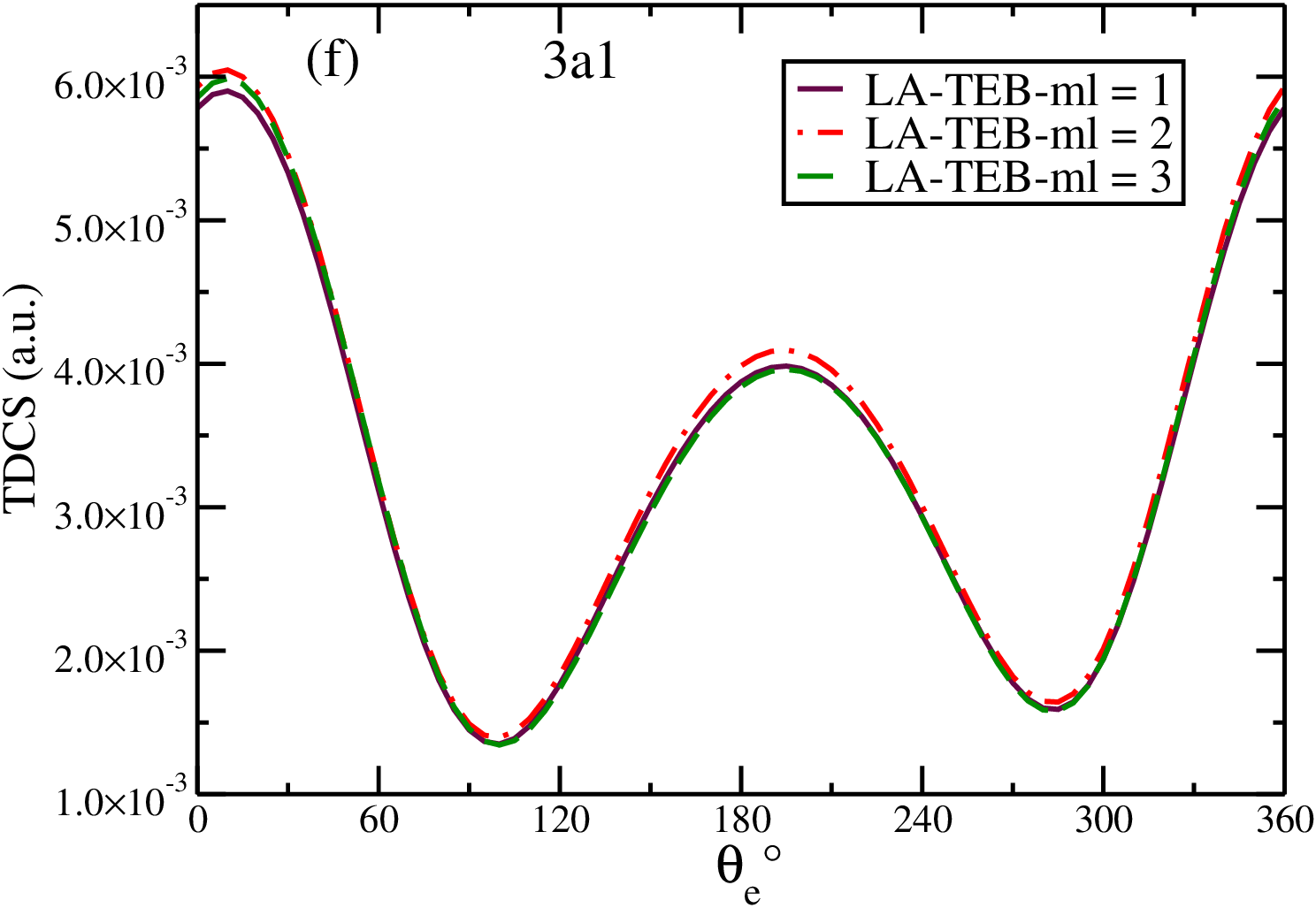}
					
					\end{tabular}
				
				\begin{tabular}{cc}
					\includegraphics[width=8.00cm]{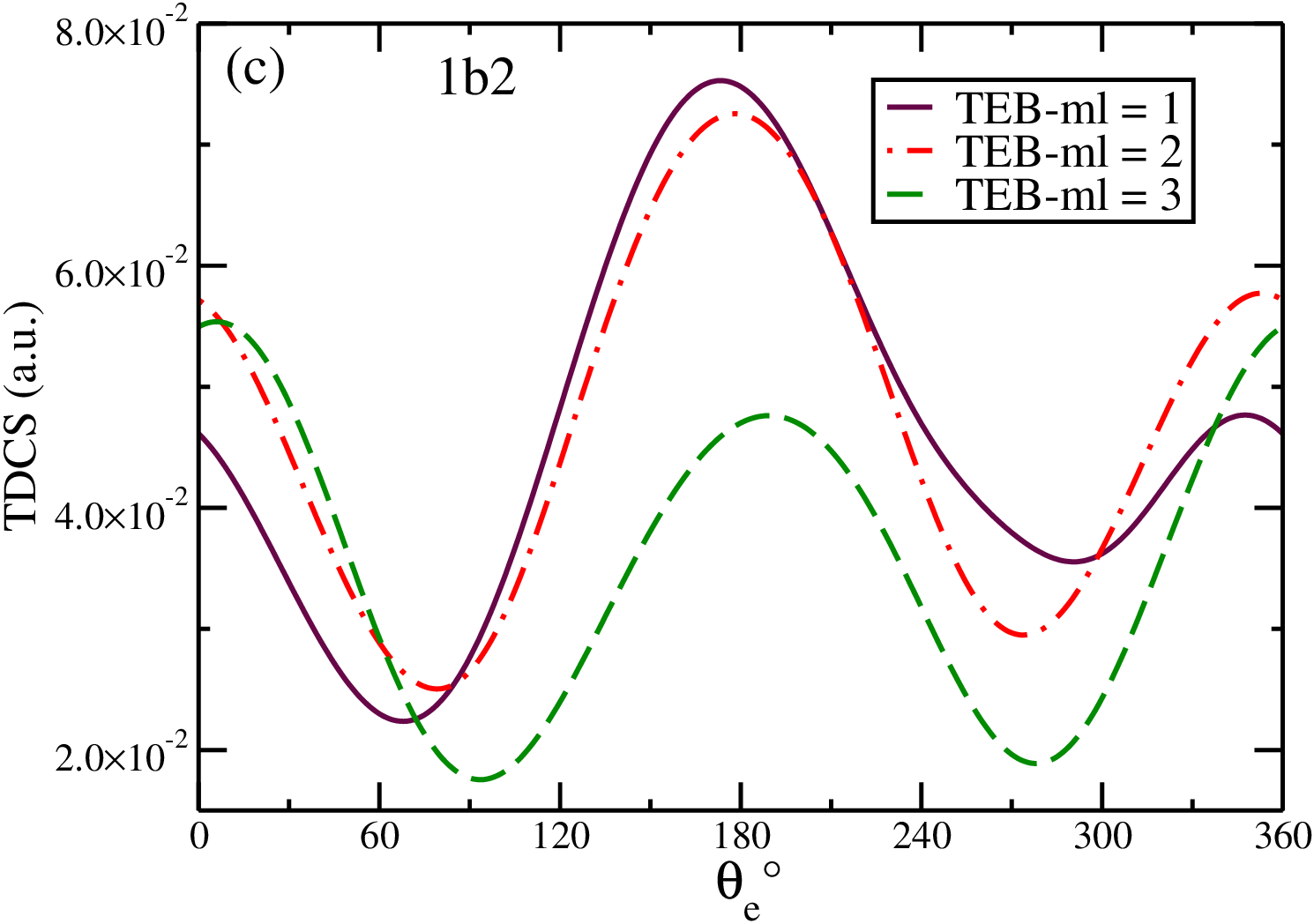}\ & \includegraphics[width=8.00cm]{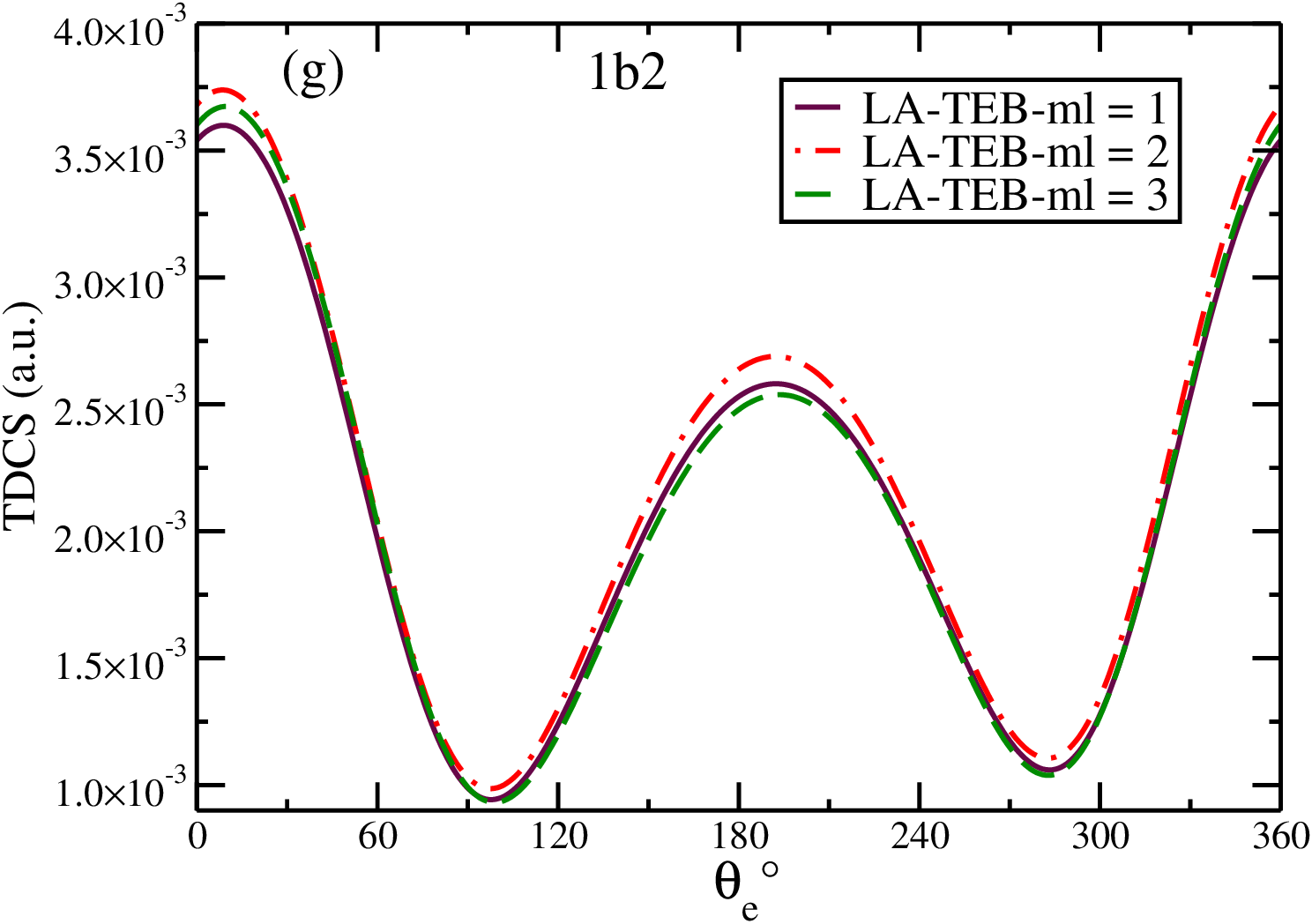}
					
				\end{tabular}
                \begin{tabular}{cc}
					\includegraphics[width=8.00cm]{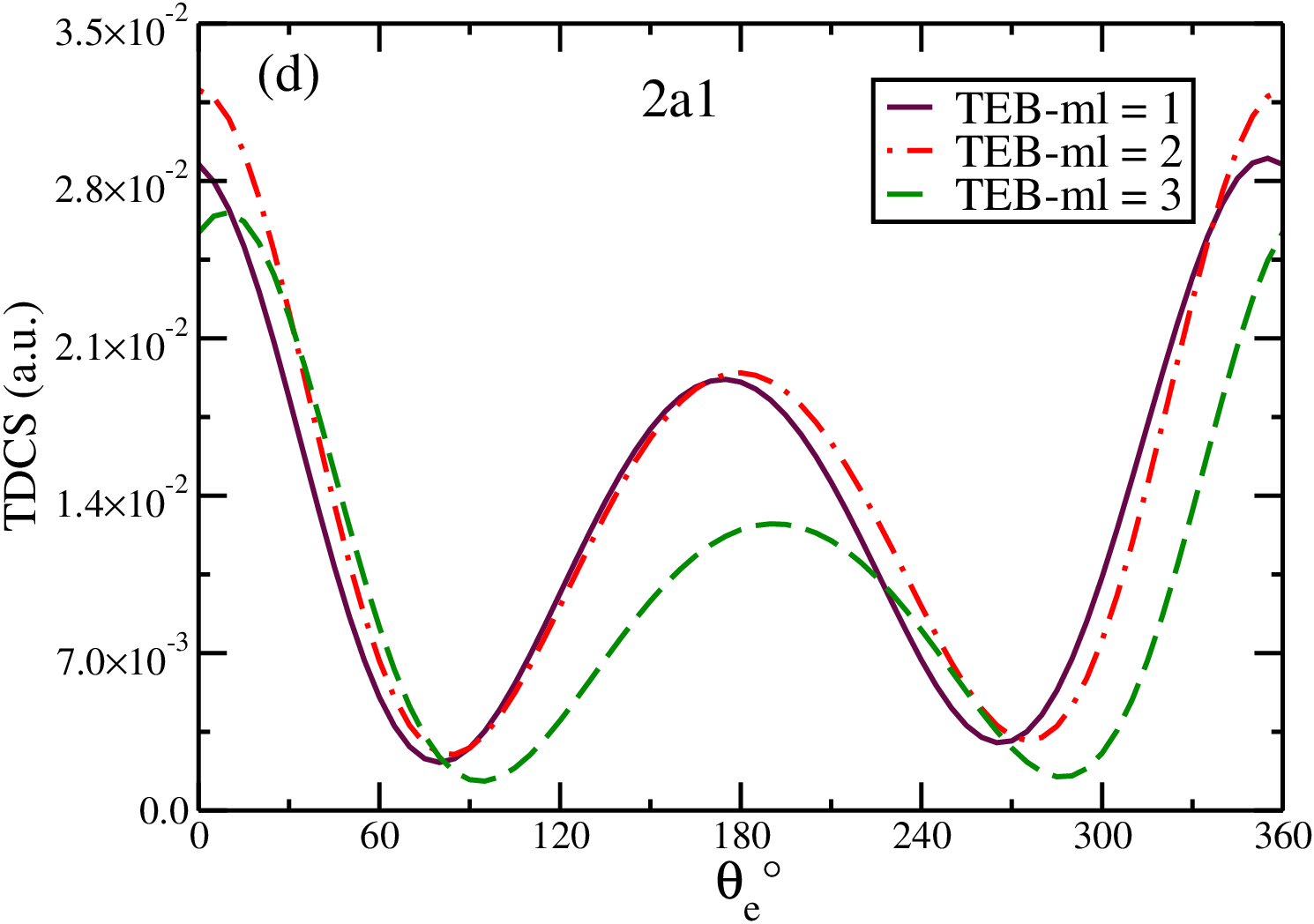}\ & \includegraphics[width=8.00cm]{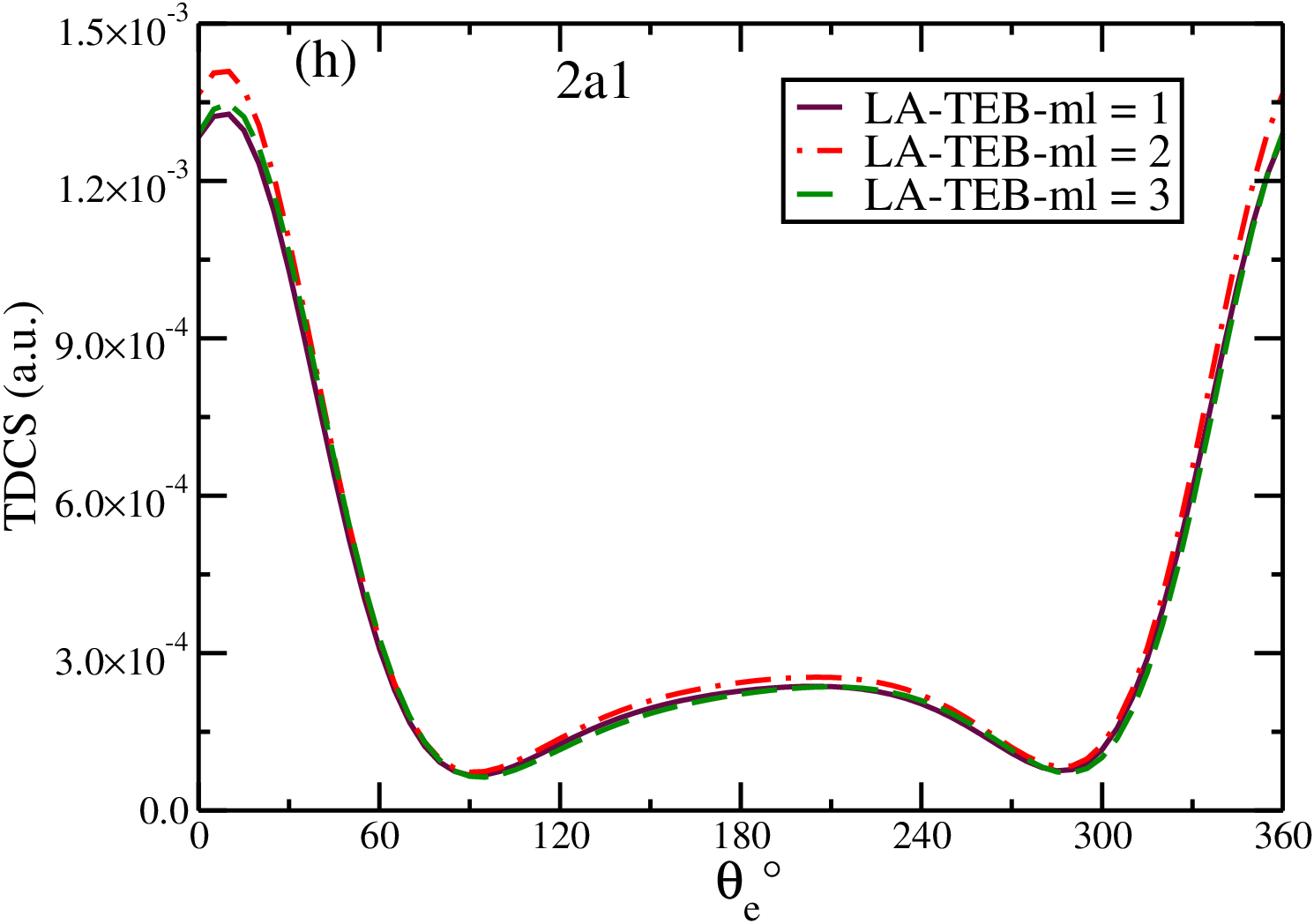}
				
				\end{tabular}

				\caption{Same as Fig \ref{fig:2} except with impact parameter $\textbf{b}$ = 0.1 nm.}   
			\label{fig:7}
			\end{figure*}
Finally, we investigate the angular profile of the TDCS for TEB (left column in Fig \ref{fig:7}) and LA-TEB (right column in Fig \ref{fig:7}) for the impact parameter of \textbf{b} = 0.1 nm to study the effect of nonzero impact parameter on laser-assisted (e,2e) processes by twisted electrons. Calculations have been performed for OAM numbers $m_l$ = 1 (maroon solid curve), 2 (red dashed-dotted curve), and 3 (green dashed curve). For $p-like$ orbitals, we observed a two-peak structure: a forward peak near $\theta_e$ = 0$^\circ$ (or 360$^\circ$) and a backward peak at $\theta_e$ = 180$^\circ$. The backward peak at $\theta_e$ = 180$^\circ$ is more prominent than the forward peaks (see Figs \ref{fig:7} (a), (b) and (c)). We observed that as the OAM number $m_l$ increases from $m_l$ = 1 (solid maroon curve in Figs \ref{fig:7} (a), (b), (c)) to $m_l$ = 3 (green dashed curve in Figs \ref{fig:7} (a), (b), (c)) the magnitude of TDCS decreases. For orbital $2a_1$ with $s-like$ (Fig \ref{fig:7}(d)), we observed two peak structure: a forward peak (near $\theta_e$ = 0$^\circ$ or 360$^\circ$) and a backward peak at $\theta_e$ = 180$^\circ$. Unlike the $p-like$ orbitals for $2a_1$ orbital, the forward peaks are more prominent (see Figs \ref{fig:7} (a), (b), (c) and (d)). 
   In laser-assisted calculations (see Fig \ref{fig:7} right column) for $p-like$ orbitals, we observed dual-peak structure, a forward peak $\theta_e$ = 0$^\circ$ (or 360$^\circ$) and a backward peak at $\theta_e$ = 180$^\circ$. But unlike later cases, for these orbitals, forward peaks dominate over backward (see Figs \ref{fig:7} (e), (f) and (g)). In the laser-assisted calculations the magnitude of TDCS is higher for even values of OAM ($m_l$ = 2) as compare to that of odd values of $m_l$ = 1 and 3 (see red curve in Figs. \ref{fig:7}(e), (f), and (g)). However, in the absence of a laser field, the TDCS magnitude decreases with increasing OAM, as observed in Figs. \ref{fig:7}(a), (b), and (c). For the
 $2a_1$ orbital, a two-peak structure is observed, similar to $p-like$ orbitals, a forward peak near $\theta_e$ = 0$^\circ$ or 360$^\circ$ and a backward peak at $\theta_e$ = 180$^\circ$. But unlike $p-like$ orbitals, the ratio of forward to backward peak gets enhanced for $2a_1$ orbital (see Fig \ref{fig:7}(h)).

 		\section{Conclusion}
		\label{conc}
			
			This paper presents a theoretical study on the laser-assisted (e,2e) process for both conventional plane and twisted electron beams on $H_{2}O$ molecule. Our theoretical model for the (e,2e) process is formulated for a linearly polarised laser field in the first-Born approximation. The incident projectile is described by Volkov wave function, and the slow-moving ejected electron by Coulomb-Volkov wave function. Calculations for the laser-assisted TDCS have been performed for different orientations of the laser field. The angular distributions of TDCS have been studied for laser field polarization parallel to the incident momentum ($\varepsilon_0$ $\parallel$ $k_i$), parallel to momentum transfer ($\varepsilon_0$ $\parallel$ $\Delta$) and perpendicular to the momentum transfer ($\varepsilon_0$ $\perp$ $\Delta$). 
            It was observed that the laser field significantly modifies the angular distribution of TDCS. For the two orientations 
            $\varepsilon_0$ $\parallel$ $k_i$ and $\varepsilon_0$ $\perp$ $\Delta$, we observed oscillatory nature of TDCS but for the orientation $\varepsilon_0$ $\parallel$ $\Delta$ we observed only recoil peak for $p-like$ character orbitals and whereas dual peak; a recoil and binary peak for the $s-like$ character orbital. Out of the three orientations of the laser field employed in this study, the orientation $\varepsilon_0$ $\parallel$ $\Delta$ has the highest magnitude of TDCS compared to the other two cases ($\varepsilon_0$ $\parallel$ $k_i$) and ($\varepsilon_0$ $\perp$ $\Delta$). 
            We also investigate the influence of TEB parameters on the angular distribution of the TDCS in the presence of a laser field. In our results of TEB with and without laser field, we observed a two-peak structure: a forward and backward peak in the direction $\theta_e$ = 0$^\circ$ or 360$^\circ$ and $\theta_e$ = 180$^\circ$ respectively. However, for LA-TEB for the orbitals $1b_1$, $3a_1$, and $1b_2$ with $p-like$ character, the angular distribution of TDCS is more symmetric than TEB. 
            For the orbital $2a_1$ with $s-like$ character, we observed that the laser field effects are more prominent in the vicinity of the backward peak; as the backward peaks get suppressed as compared to TEB.
            For the laser-assisted processes (LA-TEB), the difference in magnitude of TDCS for different values of $m_l$ is smaller than that of the TEB. In case of TEB studies, the presence of a laser field affects angular distribution of TDCS more for nonzero impact parameter as compare to the zero impact parameter. 
     In this study, we also observed that the presence of the laser field more dramatically affects the angular profile of TDCS for plane wave as compared to that for the TEB.

            This manuscript presents our first attempt to examine the laser-assisted (e,2e) process on a molecular target using both plane-wave and twisted electron beams. In the future, one can extend this study further by exploring the effects of additional laser parameters, such as frequency, different polarization states (including elliptical and circular polarization), and laser field strength.
           Our theoretical model uses the 1CW to study the TDCS. In the future, one can use more sophisticated models, such as DWBA, 2CW, BBK, and DS3C \cite{champion2006single,ren2017electron,singh2019low} for better insight into these processes.

			\nocite{apsrev41Control}
			\bibliography{MS_CRSV2}  
		\end{document}